\documentclass[aps,twocolumn,prd,showpacs,superscriptaddress]{revtex4}
\usepackage[dvips]{graphicx}
\usepackage{amsmath}
\usepackage{amssymb}
\newcommand{\ud}{\mathrm{d}}
\newcommand{\dirac}{\partial\llap{$\diagup$\kern-2pt}}

\newcommand{\diag}{\mathrm{diag}}

\newcommand{\e}{\mathrm{e}}
\newcommand{\be}{\begin{equation}}
\newcommand{\ee}{\end{equation}}
\newcommand{\bea}{\begin{eqnarray}}
\newcommand{\eea}{\end{eqnarray}}
\newcommand{\bsub}{\begin{subequations}}
\newcommand{\esub}{\end{subequations}}
\begin{document}

\title{The phase diagram of neutral quark matter: \\
The effect of neutrino trapping}


\author{Stefan B.\ R\"uster}
\email{ruester@th.physik.uni-frankfurt.de}
\affiliation{Institut f\"ur Theoretische Physik, 
J.W.\ Goethe-Universit\"at,
D-60438 Frankfurt am Main, Germany}

\author{Verena Werth}
\email{verena.werth@physik.tu-darmstadt.de}
\affiliation{Institut f\"ur Kernphysik, 
Technische Universit\"at Darmstadt,
D-64289 Darmstadt, Germany}

\author{Michael Buballa}
\email{michael.buballa@physik.tu-darmstadt.de}
\affiliation{Institut f\"ur Kernphysik, 
Technische Universit\"at Darmstadt,
D-64289 Darmstadt, Germany}

\author{Igor A.\ Shovkovy}
\email{shovkovy@th.physik.uni-frankfurt.de}
  \altaffiliation[on leave from ]{%
       Bogolyubov Institute for Theoretical Physics,
       03143, Kiev, Ukraine}
\affiliation{
Frankfurt Institute for Advanced Studies, J.W.\ Goethe-Universit\"at,
D-60438 Frankfurt am Main, Germany}

\author{Dirk H.\ Rischke}
\email{drischke@th.physik.uni-frankfurt.de}
\affiliation{Institut f\"ur Theoretische Physik, 
J.W.\ Goethe-Universit\"at,
D-60438 Frankfurt am Main, Germany}


\date{\today}

\begin{abstract}
We study the effect of neutrino trapping on the phase diagram 
of dense, locally neutral three-flavor quark matter within the 
framework of a Nambu--Jona-Lasinio model. In the analysis, dynamically 
generated quark masses are taken into account self-consistently. 
The phase diagrams in the plane of temperature and quark chemical 
potential, as well as in the plane of temperature and lepton-number
chemical potential are presented. We show that neutrino trapping 
favors two-flavor color superconductivity and disfavors the
color-flavor-locked phase at intermediate densities of matter. 
At the same time, the location of the critical line separating 
the two-flavor color-superconducting phase and the normal phase 
of quark matter is little affected by the presence of neutrinos. 
The implications of these results for the evolution of protoneutron 
stars are briefly discussed. 
\end{abstract}

\maketitle

\section{introduction}

Recently, interest in the structure of the phase diagram 
of neutral dense quark matter has increased considerably 
\cite{phase-d,phase-d1,phase-d2,kyoto,pd-mass,pd-other}. 
This was driven in part by several key observations 
regarding the influence of charge neutrality and $\beta$ 
equilibrium on the competition of various phases in quark 
matter \cite{absence2sc,SRP,mei,g2SC,gCFL,dSC}. 

In general, dense quark matter is expected to be color-superconducting.
(For reviews on color superconductivity, see Ref.~\cite{reviews}.)
At extremely large densities, namely when the quark chemical potential 
$\mu$ is much larger than the constituent, medium-modified quark masses, 
the ground state of matter is given by the so-called color-flavor-locked 
(CFL) phase \cite{cfl} (for studies of QCD at large densities, see also 
Ref.~\cite{weakCFL}). At the highest densities existing in stars, 
however, the chemical potential is unlikely to be much larger than 
$500$~MeV, while the constituent mass of the strange quark is not
smaller than the current mass, which is about $100$~MeV. 

In stellar matter, therefore, the heavier strange quark may not 
be able to participate in diquark Cooper pairing as easily as the 
light up and down quarks. Then, the pairing of light quarks can lead to 
the two-flavor color-superconducting (2SC) ground state \cite{cs,weak}. 
It should be pointed out, however, that the 2SC phase is subject 
to large penalties after imposing the charge neutrality 
and the $\beta$-equilibrium conditions 
\cite{absence2sc}. Indeed, when the fraction of strange quarks and 
leptons is small, the electric neutrality requires roughly twice as 
many down quarks as up quarks. In this case, Cooper pairing of up 
quarks with down quarks of opposite momenta becomes rather difficult. 
Then, depending on the details of the interaction, the gapless CFL 
(gCFL) phase \cite{gCFL}, the gapless 2SC (g2SC) phase \cite{g2SC}, 
or even the normal quark (NQ) matter phase may be more favored. 
Also, there exist other reasonable possibilities 
\cite{mix,spin-1,cryst,meson-cond,hong,gluonic} 
which, in view of the known instabilities in the gapless phases 
\cite{instability}, are considered to be very promising. 

In Ref.~\cite{pd-mass}, we obtained the phase diagram of neutral 
quark matter, described by the NJL-type model with the parameter 
set from Ref.~\cite{RKH}. In contrast to previous studies in 
Refs.~\cite{phase-d,phase-d1,phase-d2}, the dynamically generated 
quark masses were treated self-consistently in Ref.~\cite{pd-mass}.
(For earlier studies on color superconductivity, treating quark 
masses as dynamical quantities, see also Refs.~\cite{BO,HZC}.) In 
the present article we follow the same approach to study the effect 
of a nonzero neutrino (or, more precisely, lepton-number) chemical 
potential on the structure of the phase diagram. This is expected 
to have a potential relevance for the physics of protoneutron stars 
where neutrinos are trapped during the first few seconds of the 
stellar evolution.

The effect of neutrino trapping on color-superconducting quark matter
has been previously discussed in Ref.~\cite{SRP}. It was found that
a nonzero neutrino chemical potential favors the 2SC phase and 
disfavors the CFL phase. This is not unexpected because the 
neutrino chemical potential is related to the conserved lepton 
number in the system and therefore it also favors the presence 
of (negatively) charged leptons. This helps 2SC-type pairing
because electrical neutrality in quark matter can be achieved 
without inducing a very large mismatch between the Fermi surfaces 
of up and down quarks. The CFL phase, on the other hand, is 
electrically and color neutral {\it in the absence} of charged 
leptons when $T =0$ \cite{enforced}. A nonzero neutrino 
chemical potential can only spoil CFL-type pairing.

In the present paper we extend the analysis of Ref.~\cite{SRP} 
by performing a more systematic survey of the phase diagram in 
the space of temperature, quark and lepton-number chemical 
potentials. This also includes the possibility of gapless phases, 
which have not been taken into account in Ref.~\cite{SRP}. 
While such phases are generally unstable at zero temperature 
\cite{instability}, this is not always the case at nonzero temperature 
\cite{F-proc}. Keeping this in mind, we shall merely localize 
the ``problematic" regions in the phase diagram, where unconventional 
pairing is unavoidable. We shall refrain, however, from speculating 
on various possibilities for the true ground state (see, e.g., 
\cite{mix,spin-1,cryst,meson-cond,hong,gluonic}), since these 
are still under debate.

In the application to protoneutron stars, it is of interest to 
cover a range of parameters that could provide a total lepton 
fraction in quark matter of up to about 0.4. This is the value 
of the lepton-to-baryon charge ratio in iron cores of progenitor 
stars. Because of the conservation of both lepton and baryon 
charges, this value is also close to the lepton fraction in 
protoneutron stars at early times, when the leptons did not 
have a chance to diffuse through dense matter and escape from 
the star \cite{prakash-et-al}. The densities in the central 
regions of protoneutron stars can reach values of up to about 
10 times larger than the nuclear saturation density $\rho_0 
\approx 0.15~\mbox{fm}^{-3}$ \cite{prakash-et-al}. In the 
model studied in the present paper, almost the whole range of 
possibilities shall be covered by restricting the quark chemical 
potential to $\mu\alt 500~\mbox{MeV}$ and the neutrino chemical 
potential to $\mu_{\nu_{e}}\alt 400~\mbox{MeV}$.

This paper is organized as follows. In Sec.~\ref{eos}, we 
briefly introduce the model. The general effects of neutrino 
trapping on two- and three-flavor quark matter are discussed 
in Sec.~\ref{simple}. The results for the phase diagram of 
quark matter with neutrino trapping are presented in 
Sec.~\ref{results}. In Sec.~\ref{conclusions}, we summarize 
our results and discuss their implications for the evolution of 
protoneutron stars.

\section{Model}
\label{eos}

We consider a system of up, down, and strange quarks, in weak 
equilibrium with charged leptons and the corresponding neutrinos. 
We assume that the lepton sector of the model is given by an 
ideal gas of massive electrons ($m_e \approx 0.511$~MeV) and 
muons ($m_\mu \approx 105.66$~MeV), as well as massless electron 
and muon neutrinos. We do not take into account the $\tau$ lepton,
which is too heavy to play any role in dense matter. Also, we 
neglect the possibility of neutrino mixing, and therefore we do 
not take into account the $\tau$ neutrino either. 

In the quark sector, we use the same three-flavor Nambu--Jona-Lasinio 
model as in Ref.~\cite{pd-mass}. The Lagrangian density reads 
\begin{eqnarray}
\mathcal{L} &=& \bar \psi \, ( i \dirac - \hat{m} \, ) \psi 
+G_S \sum_{a=0}^8 \left[ \left( \bar \psi \lambda_a \psi \right)^2 
+ \left( \bar \psi i \gamma_5 \lambda_a \psi \right)^2 \right] 
\nonumber \\
&+& G_D \sum_{\gamma,c} \left[\bar{\psi}_{\alpha}^{a} i \gamma_5
\epsilon^{\alpha \beta \gamma}
\epsilon_{abc} (\psi_C)_{\beta}^{b} \right] \left[ 
(\bar{\psi}_C)_{\rho}^{r} i \gamma_5
\epsilon^{\rho \sigma \gamma} \epsilon_{rsc} \psi_{\sigma}^{s} 
\right] 
\nonumber \\
&-& K \left\{ \det_{f}\left[ \bar \psi \left( 1 + \gamma_5 \right) \psi
\right] + \det_{f}\left[ \bar \psi \left( 1 - \gamma_5 \right) \psi
\right] \right\} \;,
\label{Lagrangian}
\end{eqnarray}
where the quark spinor field $\psi_{\alpha}^{a}$ carries color 
($a=r,g,b$) and flavor ($\alpha=u,d,s$) indices. 
$\hat{m} = \diag_{f}(m_u, m_d, m_s)$ is the matrix of quark current 
masses; $\lambda_a$ ($a=1,\ldots,8$) are the Gell-Mann matrices in 
flavor space, and $\lambda_0\equiv \sqrt{2/3}\,\openone_{f}$. The charge 
conjugate spinor is defined as $\psi_C = C \bar \psi^T$ and $\bar
\psi_C = \psi^T C$, where $\bar\psi=\psi^\dagger \gamma^0$ is the 
Dirac conjugate spinor and $C=i\gamma^2 \gamma^0$ is the charge 
conjugation matrix.

In the model at hand, there are six mutually commuting conserved 
charge densities. They split naturally into the following three 
classes.

\begin{itemize}

\item[(i)] {\em Electric charge.} This is related to the $U(1)$ 
symmetry of electromagnetism. The corresponding charge 
density is given by
\be
   n_Q = \langle\psi^\dagger Q \psi\rangle - n_e - n_\mu~,
\ee
where $Q=\mbox{diag}_{f}(\frac23,-\frac13,-\frac13)$ is the electric 
charge matrix of the quarks, and $n_e$ and $n_\mu$ denote the number 
densities of electrons and muons, respectively.

\item[(ii)] {\em Two lepton charges}. As long as the neutrinos 
are trapped and their oscillations are neglected, the lepton 
family numbers are conserved. The corresponding densities 
read
\be
   n_{L_e} = n_e + n_{\nu_e}~, \qquad 
   n_{L_\mu} = n_\mu + n_{\nu_\mu}~.
\ee

\item[(iii)] {\em Baryon number and two color charges}.
The $SU(3)$-color symmetry and $U(1)$ baryon number 
symmetry imply the conservation of three independent charge 
densities in the quark sector,
\be
   n = \langle\psi^\dagger\psi\rangle, \quad
   n_3 = \langle\psi^\dagger T_3 \psi\rangle, \quad
   n_8 = \langle\psi^\dagger T_8 \psi\rangle, 
\ee
where $T_3=\mbox{diag}_c(\frac12,-\frac12,0)$ and 
      $\sqrt{3}T_8=\mbox{diag}_c(\frac12,\frac12,-1)$
are the matrices associated with the two mutually commuting 
color charges of the $SU(3)_c$ gauge group. Note that $n$ 
is the quark number density that is related to the baryon 
number density as follows: $n_B=\frac13 n$. An alternative 
choice of the three conserved charges is given by the number 
densities of red, green, and blue quarks, i.e., 
$n_{r}=n_B+n_3+n_8/\sqrt{3}$, $n_{g}=n_B-n_3+n_8/\sqrt{3}$, 
and $n_{b}=n_B-2n_8/\sqrt{3}$.

\end{itemize}
The six conserved charge densities defined above are related to the six
chemical potentials of the model. These are the quark chemical potential 
$\mu = \mu_B/3$, the two color chemical potentials $\mu_3$ and $\mu_8$, 
the electric charge chemical potential $\mu_Q$, and the two lepton-number 
chemical potentials $\mu_{L_e}$ and $\mu_{L_\mu}$.

In chemical equilibrium, the chemical potentials of all individual 
quark and lepton species can be expressed in terms of these six 
chemical potentials according to their content of conserved charges.
For the quarks, which carry quark number, color, and electric charge,
this is related to the diagonal elements of the following matrix:
\begin{equation}
\mu_{ab}^{\alpha\beta} = \left(
  \mu \delta^{\alpha\beta} 
+ \mu_Q Q^{\alpha\beta} \right)\delta_{ab} 
+ \left[ \mu_3 \left(T_3\right)_{ab} 
+ \mu_8 \left(T_8\right)_{ab} \right] \delta^{\alpha\beta}~,
\label{mu-f-i}
\end{equation}
i.e., the chemical potential of a quark with flavor $\alpha$ and 
color $a$ is given by $\mu_a^\alpha \equiv \mu_{aa}^{\alpha\alpha}$. 
The neutrinos, on the other hand, carry lepton number only. Then, 
\be
   \mu_{\nu_e} = \mu_{L_e}~, \qquad    
   \mu_{\nu_\mu} = \mu_{L_\mu}~.
\ee
Finally, electrons and muons carry both, lepton number and electric
charge, hence
\be
   \mu_e  = \mu_{L_e} - \mu_Q~, \qquad
   \mu_\mu  = \mu_{L_\mu} - \mu_Q~.
\ee
As in Ref.~\cite{pd-mass}, the quark part of our model is treated in
the mean-field (Hartree) approximation, allowing for the presence of 
both quark-antiquark condensates and scalar diquark condensates. In
the model, the constituent quark masses are defined by the following 
relation:
\begin{equation}
M_\alpha = m_\alpha - 4 G_S \sigma_\alpha 
+ 2 K \sigma_\beta \sigma_\gamma \; ,
\label{Mi}
\end{equation}
where $\sigma_\alpha =\langle\bar\psi_\alpha^a\psi_\alpha^a\rangle$ 
(no sum over $\alpha$) are the quark-antiquark condensates, and 
the set of indices $(\alpha, \beta, \gamma)$ is a permutation of 
$(u,d,s)$.

Diquark pairing in the model is described by three gap parameters
\be
\Delta_{c} = 2 G_D\langle(\bar{\psi}_C)_{\alpha}^{a} i \gamma_5 
\epsilon^{\alpha \beta c} \epsilon_{a b c} \psi_{\beta}^{b}\rangle
\ee
(no sum over $c$). By definition, the parameters $\Delta_1$, $\Delta_2$, 
and $\Delta_3$ correspond to the down-strange, up-strange, and up-down 
diquark condensates, respectively. 

By following the same steps in the derivation as in Ref.~\cite{pd-mass}
and including the ideal-gas contributions for the leptons, we arrive 
at the following expression for the pressure:
\begin{eqnarray}
p &=& \frac{1}{2 \pi^2} \sum_{i=1}^{18} \int_0^\Lambda \ud k \, k^2
\left[ |\epsilon_i| + 2 T \ln \left( 1 + \e^{-
\frac{|\epsilon_i|}{T}} \right) \right] \nonumber \\
&+& 4 K \sigma_u \sigma_d \sigma_s
- \frac{1}{4 G_D} \sum_{c=1}^{3} \left| \Delta_c \right|^2
-2 G_S \sum_{\alpha=1}^{3} \sigma_{\alpha}^2
\nonumber \\
&+& \sum_{l=e,\mu} \left[ \frac{T}{\pi^2} \sum_{\epsilon=\pm}
\int_0^\infty \ud k \, k^2
\ln \left( 1 + \e^{-\frac{E_l-\epsilon\mu_l}{T}}\right)
\right. \nonumber \\
&+& \left. \frac{1}{24\pi^2} \left( \mu_{\nu_l}^4 +
2 \pi^2 \mu_{\nu_l}^2 T^2 + \frac{7}{15} \pi^4
T^4 \right) \right] \; ,
\label{pressure}
\end{eqnarray}
where $\epsilon_i$ are eighteen independent positive-energy eigenvalues, 
see Ref.~\cite{pd-mass} for details. The pressure in Eq.~(\ref{pressure}) 
has a physical meaning only at stationary points with respect to 
variations of the chiral and color-superconducting condensates. Thus,
$\sigma_{\alpha}$ and $\Delta_c$ must satisfy the following set of 
six gap equations:
\bsub
\label{gapeqns}
\bea
\frac{\partial p}{\partial \sigma_{\alpha}} &=& 0\; , \\
\frac{\partial p}{\partial \Delta_c} &=& 0  \; .
\eea
\esub
To enforce the conditions of local charge neutrality in dense matter,
one also requires that the following three equations are satisfied:
\bsub
\label{neutrality}
\bea
n_Q &\equiv& \frac{ \partial p }{\partial \mu_Q} = 0 \; , \\
n_3 &\equiv& \frac{ \partial p }{\partial \mu_3} = 0 \; , \\
n_8 &\equiv& \frac{ \partial p }{\partial \mu_8} = 0 \; .
\eea
\esub
We should note that, in general, the restriction to the color
charges $n_3$ and $n_8$ is not sufficient, but one has to ensure 
that the densities $n_a = \langle\psi^\dagger T_a \psi\rangle$ 
vanish for {\it all} generators $T_a$ of $SU(3)_c$, i.e., 
$a = 1, \dots, 8$~\cite{BuSh}. However, for all phases we 
are considering here, we have checked explicitly that the 
condition $n_a=0$ is automatically satisfied for $a\neq 3,8$.
Thus, we are left with Eqs.~(\ref{neutrality}). By solving 
these, we determine the values of the three corresponding chemical 
potentials $\mu_Q$, $\mu_3$, and $\mu_8$ for a given set of the other 
chemical potentials, $\mu$, $\mu_{L_e}$, $\mu_{L_\mu}$, and for a given 
temperature $T$. In general, therefore, the phase diagram of dense quark 
matter with neutrino trapping should span a four-dimensional parameter 
space.  

Note that instead of using the chemical potentials, $\mu$, $\mu_{L_e}$, 
and $\mu_{L_\mu}$ as free parameters in the study of the phase 
diagram, one may also try to utilize the quark number density and 
the two lepton fractions,
\bsub
\label{qldens}
\bea
n         &\equiv& \frac{ \partial p }{\partial \mu} \; , \\
Y_{L_e}   &\equiv& 3 \frac{n_{L_e}}{n} \; , \\
Y_{L_\mu} &\equiv& 3 \frac{n_{L_\mu}}{n} \; ,
\eea
\esub
where the two lepton densities $n_{L_i}$ are defined by
\be
n_{L_e} \equiv \frac{ \partial p }{\partial \mu_{L_e}} \; , \qquad
n_{L_\mu} \equiv \frac{ \partial p }{\partial \mu_{L_\mu}} \; .
\ee
In some cases, the choice of $n$, $Y_{L_e}$ and $Y_{L_\mu}$ as free 
parameters is indeed very useful. For instance, this is helpful in 
order to determine the initial state of matter inside the protoneutron 
star at very early times, when the lepton fractions are approximately 
the same as in the progenitor star (i.e., $Y_{L_e} \approx 0.4$ and 
$Y_{L_\mu} = 0$). The problem is, however, that such an approach becomes
ambiguous in the vicinity of first-order phase transitions, where the 
baryon number density as well as the lepton fractions are in general 
discontinuous. For this reason, it is more appropriate to study the 
phase structure of (dense) QCD at given fixed values of the chemical 
potentials $\mu$, $\mu_{L_e}$, and $\mu_{L_\mu}$. Unlike densities,
the chemical potentials change continuously when the system crosses
a boundary of a first-order phase transition. (Here it should be 
noted that, because of the long-range Coulomb interaction enforcing 
the constraints $n_Q=0$, $n_3=0$ and $n_8=0$, the chemical potentials 
$\mu_{Q}$, $\mu_{3}$, and $\mu_{8}$ may change almost discontinuously 
at a boundary of a first-order phase transition.)

\section{Simplified considerations}
\label{simple}

As mentioned in the Introduction, neutrino trapping favors the 
2SC phase and strongly disfavors the CFL phase~\cite{SRP}. This 
is a consequence of the modified $\beta$-equilibrium condition 
in the system. In this section, we would like to emphasize that 
this is a model-independent effect. In order to understand the 
physics behind it, it is instructive to start our consideration 
from a very simple toy model. Later, many of its qualitative 
features will be also observed in our self-consistent numerical 
analysis of the NJL model. 

Let us first assume that strange quarks are very heavy and
consider a gas of non-interacting massless up and down
quarks in the normal phase at $T=0$. As required by $\beta$ 
equilibrium, electrons and electron neutrinos are also 
present in the system. (Note that in this section we neglect 
muons and muon neutrinos for simplicity.) 

In the absence of Cooper pairing, the densities of quarks and 
leptons are given by
\be
  n_{u,d} \;=\; \frac{\mu_{u,d}^3}{\pi^2}~, \quad
  n_e \;=\; \frac{\mu_e^3}{3\pi^2}~, \quad
  n_{\nu_e} \;=\; \frac{\mu_{\nu_e}^3}{6\pi^2}~. 
\ee
Expressing the chemical potentials through $\mu$, $\mu_Q$ and
$\mu_{L_e}$, and imposing electric charge neutrality, one arrives 
at the following relation:
\be
    2(1 +\frac{2}{3}y)^3 \,
-\, (1 -\frac{1}{3}y)^3 \,-\,
     (x - y)^3  \;=\; 0~,
\label{toy2}
\ee
where we have introduced the chemical potential ratios 
$x = \mu_{L_e}/\mu$ and $y = \mu_Q/\mu$. The above cubic 
equation can be solved for $y$ (electric chemical
potential) at any given $x$ (lepton-number chemical
potential). The result can be used to calculate the ratio
of quark chemical potentials, $\mu_d/\mu_u = (3-y)/(3+2y)$.

\begin{figure*}[h!]
  \begin{center}
    \includegraphics[width=0.7\textwidth]{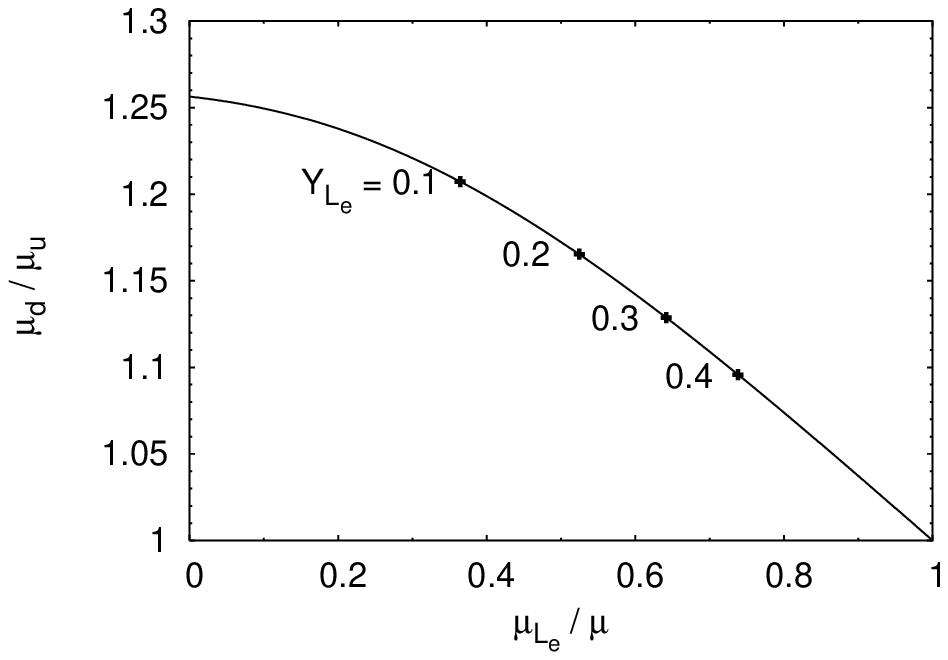}
    \caption{Ratio of down and up quark chemical potentials 
     as a function of $\mu_{L_e}/\mu$ in the toy model of
     Sec.~\ref{simple}. The crosses mark the solutions at 
     several values of the lepton fraction.}
    \label{toy}
  \end{center}
\end{figure*}

The ratio $\mu_d/\mu_u$ as a function of $\mu_{L_e}/\mu$ is 
shown in Fig.~\ref{toy}. At vanishing $\mu_{L_e}$, one finds 
$y \approx -0.219$ and, thus, $\mu_d/\mu_u \approx 1.256$ 
(note that this value is very close to $2^{1/3}\approx 1.260$). 
This result corresponds to the following ratios of the number 
densities in the system: $n_u/n_d\approx 0.504$ and $n_e/n_d
\approx 0.003$, reflecting that the density of electrons is 
tiny and the charge of the up quarks has to be balanced by 
approximately twice as many down quarks. 

At $\mu_{L_e}=\mu$, on the other hand, the real solution to 
Eq.~(\ref{toy2}) is $y=0$, i.e., the up and down Fermi momenta 
become equal. This can be seen most easily if one inverts the 
problem and solves Eq.~(\ref{toy2}) for $x$ at given $y$. When 
$y=0$ one finds $x=1$, meaning that $\mu_d=\mu_u$ and, in turn, 
suggesting that pairing between up and down quarks is unobstructed 
at $\mu_{L_e}=\mu$. This is in contrast to the case of vanishing 
$\mu_{L_e}$, when the two Fermi surfaces are split by about 25\%, 
and pairing is difficult. 

It is appropriate to mention that many features of the above 
considerations would not change much even when Cooper pairing 
is taken into account. The reason is that the corresponding 
corrections to the quark densities are parametrically suppressed 
by a factor of order $(\Delta/\mu)^2$. 

In order to estimate the magnitude of the effect in the case
of dense matter in protoneutron stars, we indicate several 
typical values of the lepton fractions $Y_{L_e}$ in Fig.~\ref{toy}. 
As mentioned earlier, $Y_{L_e}$ is expected to be of order $0.4$ 
right after the collapse of the iron core of a progenitor star. 
According to Fig.~\ref{toy}, this corresponds to $\mu_d/\mu_u 
\approx 1.1$, i.e., while the splitting between the up and down Fermi 
surfaces does not disappear completely, it gets reduced considerably
compared to its value in the absence of trapped neutrinos. This reduction 
substantially facilitates the cross-flavor pairing of up and down quarks. 
The effect is gradually washed out during about a dozen of seconds of 
the deleptonization period when the value of $Y_{L_e}$ decreases to zero. 

The toy model is easily modified to the opposite extreme of 
three massless quark flavors. Basically, this corresponds to 
replacing Eq.~(\ref{toy2}) by
\be
    2(1 +\frac{2}{3}y)^3 \,-\, 2(1 -\frac{1}{3}y)^3 \,-\,
     (x - y)^3  \;=\; 0~.
\label{toy3} 
\ee
In the absence of neutrino trapping, $x = 0$, the only real solution 
to this equation is $y=0$, indicating that the chemical potentials 
(which also coincide with the Fermi momenta) of up, down, and strange
quarks are equal. This reflects the fact that the system with equal 
densities of up, down, and strange quarks is neutral by itself, without 
electrons. With increasing $x\propto\mu_{L_e}$, the solution requires 
a nonzero $y\propto \mu_Q$, suggesting that up-down and up-strange 
pairing becomes more difficult. To see this more clearly, we can go 
one step further in the analysis of the toy model.

Let us assume that the quarks are paired in a regular, i.e., fully 
gapped, CFL phase at $T=0$. Then, as shown in Ref.~\cite{enforced},
the quark part of the matter is automatically electrically neutral. 
Hence, if we want to keep the whole system electrically and color 
neutral, there must be no electrons. Obviously, this is easily 
realized without trapped neutrinos by setting $\mu_Q$ equal to zero. 
At non-vanishing $\mu_{L_e}$ the situation is more complicated. The 
quark part is still neutral by itself and therefore no electrons are 
admitted. Hence, the electron chemical potential $\mu_e = \mu_{L_e} 
- \mu_Q$ must vanish, and consequently $\mu_Q$ should be nonzero and 
equal to $\mu_{L_e}$. It is natural to ask what should be the values of
the color chemical potentials $\mu_3$ and $\mu_8$ in the CFL phase when 
$\mu_{L_e}\neq 0$. 

In order to analyze the stress on the CFL phase due to nonzero 
$\mu_{L_e}$, we follow the same approach as in 
Refs.~\cite{absence2sc,gCFL}. In this analytical consideration, 
we also account for the effect of the strange quark mass simply by 
shifting the strange quark chemical potential by $-M_s^2/(2\mu)$. 
In our notation, pairing of CFL-type requires the following 
``common'' values of the Fermi momenta of paired quarks:
\begin{subequations}
\bea
p_{F,(ru,gd,bs)}^{\rm common} &=& \mu -\frac{M_s^2}{6\mu},\\
p_{F,(rd,gu)}^{\rm common} &=& 
    \mu + \frac{\mu_{Q}}{6} + \frac{\mu_{8}}{2\sqrt{3}},\\
p_{F,(rs,bu)}^{\rm common} &=& 
    \mu +\frac{\mu_{Q}}{6} + \frac{\mu_{3}}{4}
   - \frac{\mu_{8}}{4\sqrt{3}} - \frac{M_s^2}{4\mu},\\
p_{F,(gs,bd)}^{\rm common} &=& 
    \mu -\frac{\mu_{Q}}{3} - \frac{\mu_{3}}{4}
   - \frac{\mu_{8}}{4\sqrt{3}} - \frac{M_s^2}{4\mu}.
\eea
\end{subequations}
These are used to calculate the pressure in the toy model,
\bea
p^{\rm (toy)} &=& \frac{1}{\pi^2} \sum_{a=1}^{3}\sum_{\alpha=1}^{3} 
\int_{0}^{p_{F,a\alpha}^{\rm common}}\left(\mu_a^{\alpha}-p\right)
p^2 dp \nonumber\\
&&+3\frac{\mu^2\Delta^2}{\pi^2}
+\frac{(\mu_{L_e}-\mu_Q)^4}{12\pi^2}+\frac{\mu_{L_e}^4}{24\pi^2}.
\eea
By making use of this expression, one easily derives the neutrality 
conditions as in Eq.~(\ref{neutrality}). In order to solve them, it 
is useful to note that
\be
n_{Q}-n_{3}-\frac{1}{\sqrt{3}}n_{8} = \frac{(\mu_Q-\mu_{L_e})^3}{3\pi^2}.
\ee
Thus, it becomes obvious that charge neutrality requires 
$\mu_Q=\mu_{L_e}$. The other useful observation is that the 
expression for $n_{3}$ is proportional to $\mu_3+\mu_Q$. So,
it is vanishing if (and only if) $\mu_3=-\mu_Q$, which means 
that $\mu_3=-\mu_{L_e}$. Finally, one can check that the third 
neutrality condition $n_{8}=0$ requires
\be
\mu_{8} = -\frac{\mu_{L_e}}{\sqrt{3}}-\frac{M_{s}^{2}}{\sqrt{3}\mu}.
\ee
The results for the charge chemical potentials $\mu_Q$, $\mu_3$,
and $\mu_{8}$ imply the following magnitude of stress on pairing
in the CFL phase:
\begin{subequations}
\bea
\delta\mu_{(rd,gu)} &=& \frac{\mu_{g}^{u}-\mu_{r}^{d}}{2} 
                     = \mu_{L_e},\\
\delta\mu_{(rs,bu)} &=& \frac{\mu_{b}^{u}-\mu_{r}^{s}}{2}  
                     = \mu_{L_e}+\frac{M_s^2}{2\mu},\\
\delta\mu_{(gs,bd)} &=& \frac{\mu_{b}^{d}-\mu_{g}^{s}}{2}  
                     = \frac{M_s^2}{2\mu}.
\eea
\label{mismatch}
\end{subequations}
Note that there is no mismatch between the values of the chemical 
potentials of the other three quarks, $\mu_{r}^{u}=\mu_{g}^{d}=
\mu_{b}^{s}=\mu-M_s^2/(6\mu)$. 

{From} Eq.~(\ref{mismatch}) we see that the largest mismatch occurs 
in the $(rs,bu)$ pair (for positive $\mu_{L_e}$). The CFL phase 
can withstand the stress only if the value of $\delta\mu_{(rs,bu)}$ 
is less than $\Delta_2$. A larger mismatch should drive a transition 
to a gapless phase exactly as in Refs.~\cite{g2SC,gCFL}. Thus,
the critical value of the lepton-number chemical potential is
\be
\mu_{L_e}^{\rm (cr)} \approx \Delta_2-\frac{M_s^2}{2\mu}.
\label{mu_L^cr}
\ee
When $\mu_{L_e}>\mu_{L_e}^{\rm (cr)}$, the CFL phase turns into 
the gCFL$^\prime$ phase, which is a variant of the gCFL phase 
\cite{gCFL}. By definition, the gapless mode with a linear 
dispersion relation in the gCFL$^\prime$ phase is $rs$--$bu$ 
instead of $gs$--$bd$ as in the standard gCFL phase. 
(Let us remind that the mode $a\alpha$--$b\beta$ is defined 
by its dispersion relation which interpolates between the 
dispersion relations of hole-type excitations of $a\alpha$-quark 
at small momenta, $k\ll \mu_{a}^{\alpha}$, and particle-type 
excitations of $b\beta$-quark at large momenta, 
$k\gg \mu_{b}^{\beta}$.)

In order to see what this means for the physics of protoneutron stars,
we should again try to relate the value of $\mu_{L_e}$ to the lepton 
fraction. As we have seen, there are no electrons in the (regular)
CFL phase at $T=0$. Therefore the entire lepton number is carried 
by neutrinos. For the baryon density we may neglect the pairing 
effects to first approximation and employ the ideal-gas relations.  
This yields
\be
Y_{L_e} \approx \frac{1}{6} \left(\frac{\mu_{L_e}}{\mu}\right)^3~.
\ee
Inserting typical numbers, $\mu = 500$~MeV and $\mu_{L_e} \lesssim 
\Delta \approx 50 - 100$~MeV, one finds $Y_{L_e} \lesssim 10^{-4} - 
10^{-3}$. Thus, there is practically no chance to find a sizable 
amount of leptons in the CFL phase. The constraint gets relaxed 
slightly at nonzero temperatures and/or in the gCFL phase, but 
the lepton fraction remains rather small even then (our numerical 
results indicate that, in general, $Y_{L_e} \lesssim 0.05$ in 
the CFL phase).

\section{Results}
\label{results}

The simple toy-model considerations in the previous section give
a qualitative understanding of the effect of neutrino trapping on 
the mismatch of the quark Fermi momenta and, thus, on the pairing 
properties of two- and three-flavor quark matter. Now, we turn to 
a more detailed numerical analysis of the phase diagram in the NJL 
model defined in Sec.~\ref{eos}. 

In the numerical calculations, we use the following set of model 
parameters \cite{RKH}:
\begin{subequations}
\label{model-parameters}
\begin{eqnarray}
\Lambda &=& 602.3 \; \mathrm{MeV} \; , \label{Lambda} \\
G_S \Lambda^2 &=& 1.835 \; , \\
K \Lambda^5 &=& 12.36 \; , \\
m_s &=& 140.7 \; \mathrm{MeV} \; , \\
m_{u,d} &=& 5.5 \; \mathrm{MeV} \; .
\end{eqnarray}
\end{subequations}
The parameters are chosen to reproduce several key observables of 
vacuum QCD \cite{RKH}. In this paper, our choice for the diquark 
coupling is $G_D=\frac34 G_S$, corresponding to what is called 
``intermediate coupling'' in Ref.~\cite{pd-mass}. 

In order to obtain the phase diagram, one has to determine the ground 
state of matter for each given set of the parameters. As discussed in 
Sec.~\ref{eos}, in the case of locally neutral matter with trapped 
neutrinos, there are four parameters that should be specified: the 
temperature $T$, the quark chemical potential $\mu$ as well as the two 
lepton family chemical potentials $\mu_{L_e}$ and $\mu_{L_\mu}$. After 
these are fixed, the values of the pressure in all competing neutral 
phases of quark matter should be compared. This is determined by using
the same algorithm as in Ref.~\cite{pd-mass}. The complete set of 
equations (\ref{gapeqns}) and (\ref{neutrality}) is solved for each 
of the eight phases allowed by symmetries. Then, the corresponding 
values of the pressure are determined from Eq.~(\ref{pressure}). The 
phase with the largest pressure is the ground state. 

In the present paper, we always assume that the muon lepton-number 
chemical potential vanishes, i.e., $\mu_{L_\mu}=0$. This is expected 
to be a good approximation for matter inside a protoneutron star. Our 
analysis can thus be interpreted as an extension of the $T$--$\mu$ 
phase diagram discussed in Ref.~\cite{pd-mass} into the $\mu_{L_e}$ 
direction. Consequently, the complete phase structure requires a 
three-dimensional presentation. 

\subsection{Three-dimensional phase diagram}

The general features of the phase diagram in the three-dimensional 
space spanned by the quark chemical potential $\mu$, the lepton-number 
chemical potential $\mu_{L_e}$, and the temperature $T$ are 
depicted in Fig.~\ref{phase3d}. Because of a rather complicated 
structure of the diagram, only the four main phases ($\chi$SB, 
NQ, 2SC and CFL) are shown explicitly. Although it is not 
labeled, a thin slice of a fifth phase, the uSC phase, squeezed 
in between the 2SC and CFL phases, can also be seen. While lacking 
detailed information, the phase diagram in Fig.~\ref{phase3d} gives 
a clear overall picture. Among other things, one sees, for example, 
that the CFL phase becomes strongly disfavored with increasing 
$\mu_{L_e}$ and gets gradually replaced by the 2SC phase. 

In order to discuss the structure of the phase diagram in more detail
we proceed by showing several two-dimensional slices. These are 
obtained by keeping one of the chemical potentials, $\mu$ or $\mu_{L_e}$, 
fixed and varying the other two parameters. 

\subsection{$T$--$\mu$ phase diagram}

We begin with presenting the phase diagrams at three fixed values of the 
lepton-number chemical potential, $\mu_{L_e}=0$ (upper panel), 
$\mu_{L_e}=200$~MeV (middle panel) and $\mu_{L_e}=400$~MeV (lower panel), 
in Fig.~\ref{phasediagram_mu-T}. The first of them has already been
discussed in Ref.~\cite{pd-mass}. Here we show it again only for the
purpose of comparison with the other two diagrams. The general effect
of neutrino trapping can be understood by analyzing the similarities 
and differences between the three diagrams. Note that, in this paper, 
we use the same convention for line styles as in Ref.~\cite{pd-mass}: 
thick and thin solid lines denote first- and second-order phase 
transitions, respectively; dashed lines indicate the (dis-)appearance 
of gapless modes in different phases. 

Here it is appropriate to note that, in the same model, a schematic 
version of the $T$--$\mu$ phase diagram at $\mu_{L_e}=200$~MeV was 
first presented in Ref.~\cite{SRP}, see the right panel of Fig.~4 there. 
If one ignores the complications due to the presence of the uSC phase and 
various gapless phases, the results of Ref.~\cite{SRP} are in qualitative 
agreement with our findings here. 

In order to understand the basic characteristics of different phases 
in the phase diagrams in Fig.~\ref{phasediagram_mu-T}, we also present 
the results for the dynamical quark masses, the gap parameters, and the 
charge chemical potentials. These are plotted as functions of the quark 
chemical potential in Figs.~\ref{plot0-20-40_200} and \ref{plot0-20-40_400}, 
for two different values of the temperature in the case of 
$\mu_{L_e}=200$~MeV and $\mu_{L_e}=400$~MeV, respectively.

In each of the diagrams, there are roughly four distinct regimes. 
At low temperature and low quark chemical potential, there is 
a region in which the approximate chiral symmetry is spontaneously 
broken by large $\bar \psi \psi$-condensates. This phase is denoted 
by $\chi$SB. In this regime, quarks have relatively large 
constituent masses which are close to the vacuum values, see 
Figs.~\ref{plot0-20-40_200} and \ref{plot0-20-40_400}. Here the 
density of all quark flavors is very low and even vanishes at 
$T=0$. There is no diquark pairing in this phase. The $\chi$SB 
phase is rather insensitive to the presence of a nonzero 
lepton-number chemical potential. With increasing $\mu_{L_e}$ 
the phase boundary is only slightly shifted to lower values 
of $\mu$. This is just another manifestation of the strengthening 
of the 2SC phase due to neutrino trapping.

With increasing temperature, the $\bar \psi \psi$-condensates 
melt and the $\chi$SB phase turns into the normal 
quark matter (NQ) phase where the quark masses are 
relatively small. Because of the explicit breaking 
of the chiral symmetry by the current quark masses, 
there is no need for a phase transition between the 
two regimes. In fact, at low chemical potentials we 
find only a smooth cross-over, whereas there is a 
first-order phase transition in a limited region, 
$320 \lesssim \mu \lesssim 360$~MeV. In contrast to the 
$\chi$SB regime, the high-temperature NQ phase 
extends to arbitrary large values of $\mu$.
All these qualitative features are little affected 
by the lepton-number chemical potential. 

The third regime is located at relatively low temperatures 
but at quark chemical potentials higher than in the $\chi$SB 
phase. In this region, the masses of the up and down quarks 
have already dropped to values well below their respective
chemical potentials while the strange quark mass is still 
large, see left columns of panels in Figs.~\ref{plot0-20-40_200} 
and \ref{plot0-20-40_400}. As a consequence, up and down quarks 
are quite abundant but strange quarks are essentially absent.

It turns out that the detailed phase structure in this region is very 
sensitive to the lepton-number chemical potential. At $\mu_{L_e}=0$, 
as discussed in Ref.~\cite{pd-mass}, the pairing between up and 
down quarks is strongly hampered by the constraints of 
neutrality and $\beta$ equilibrium. As a consequence, there 
is no pairing at low temperatures, $T \lesssim 10$~MeV,
and g2SC-type pairing appears at moderate temperatures, 
$10~\mbox{MeV} \lesssim T < T_c$ with the value of $T_c$ in 
the range of several dozen MeV, see the upper panel in 
Fig.~\ref{phasediagram_mu-T}. The situation changes dramatically 
with increasing the value of the lepton-number chemical potential.
Eventually, the low-temperature region of the normal phase of quark 
matter is replaced by the (g)2SC phase (e.g., at $\mu = 400$~MeV, 
this happens at $\mu_{L_e} \simeq 110$~MeV). With $\mu_{L_e}$ 
increasing further, no qualitative changes happen in this part 
of the phase diagram, except that the area of the (g)2SC phase 
expands slightly.

Finally, the region in the phase diagram at low temperatures 
and large quark chemical potentials corresponds to phases 
in which the cross-flavor strange-nonstrange Cooper pairing 
becomes possible. In general, as the strength of pairing 
increases with the quark chemical potential, the system 
passes through regions of the gapless uSC (guSC), uSC, and 
gCFL phases and finally reaches the CFL phase. (Of course, the 
intermediate phases may not always be realized.) The effect 
of neutrino trapping, which grows with increasing the lepton-number 
chemical potential, is to push out the location of the 
strange-nonstrange pairing region to larger values of $\mu$. 
Of course, this is in agreement with the general arguments in 
Sec.~\ref{simple}. 

It is interesting to note that the growth of the strangeness 
content with increasing quark chemical potential could 
indirectly be deduced from the behavior of the electric charge 
chemical potential $\mu_{Q}$ at $T=0$, see the solid lines 
in the right panels in Figs.~\ref{plot0-20-40_200} and 
\ref{plot0-20-40_400}. The value of $\mu_{Q}$ reaches its 
minimum somewhere in a range of values of the quark chemical 
potential around $\mu\simeq 440$~MeV. 
This corresponds to the point where the strange quark chemical 
potential $\mu_s \simeq \mu - \mu_Q/3$ reaches the value of the
strange quark mass (see left panels).
Hence, there are essentially no strange quarks at lower 
values of $\mu$, and a rapidly increasing amount of strange quarks 
at higher values of $\mu$. Since the latter contribute to the
electric neutralization, this is a natural explanation for
the drop of $|\mu_{Q}|$ above this point. 

As we mentioned earlier, the presence of the lepton-number chemical 
potential $\mu_{L_e}$ leads to a change of the quark Fermi 
momenta. This change in turn affects Cooper pairing of quarks, 
facilitating the appearance of some phases and suppressing 
others. As it turns out, there is also another qualitative 
effect due to a nonzero value of $\mu_{L_e}$. In particular, 
we find several new variants of gapless phases which did not 
exist at vanishing $\mu_{L_e}$. In Figs.~\ref{phasediagram_mu-T}
and \ref{phasediagramTnu}, these are denoted by the same names, 
g2SC or gCFL, but with one or two primes added. 

We define the g2SC$^{\prime}$ as the gapless two-flavor 
color-superconducting phase in which the gapless excitations 
correspond to $rd$--$gu$ and $gd$--$ru$ modes instead of the 
usual $ru$--$gd$ and $gu$--$rd$ ones, i.e., $u$ and $d$ 
flavors are exchanged as compared to the usual g2SC phase.
The g2SC$^{\prime}$ phase becomes possible only 
when the value of $(\mu_{ru} -\mu_{gd})/2\equiv(\mu_Q+\mu_3)/2$ 
is positive and larger than $\Delta_3$. The other phases are 
defined in a similar manner. In particular, the gCFL$^{\prime}$ phase,
already introduced in Sec.~\ref{simple}, is indicated by the gapless 
$rs$--$bu$ mode, while the gCFL$^{\prime\prime}$ phase has both 
$gs$--$bd$ (as in the gCFL phase) and $rs$--$bu$ gapless modes.
Definitions of all gapless phases are summarized in 
Table~\ref{table1}.

\begin{table}[t]
\caption{\label{table1}Definitions of gapless phases.}
\begin{ruledtabular}
\begin{tabular}{lll}
Name    & \begin{tabular}{l}Gapless mode(s)\\ 
        $\epsilon(k)\sim |k-k_F^{\rm eff}|$ \end{tabular} & Diquark condensate(s)\\
\hline
g2SC                  & $ru$--$gd$, $gu$--$rd$ & $\Delta_3$ \\
g2SC$^{\prime}$       & $rd$--$gu$, $gd$--$ru$ & $\Delta_3$ \\
guSC                  & $rs$--$bu$             & $\Delta_2$, $\Delta_3$  \\
gCFL                  & $gs$--$bd$             & $\Delta_1$, $\Delta_2$, $\Delta_3$ \\
gCFL$^{\prime}$       & $rs$--$bu$             & $\Delta_1$, $\Delta_2$, $\Delta_3$\\
gCFL$^{\prime\prime}$ & $gs$--$bd$, $rs$--$bu$ & $\Delta_1$, $\Delta_2$, $\Delta_3$\\
\end{tabular}
\end{ruledtabular}
\end{table}

\subsection{Lepton fraction $Y_{L_e}$}

Our numerical results for the lepton fraction $Y_{L_e}$ are shown 
in Fig.~\ref{plot_Y_0-20-40}. The thick and thin lines correspond
to two different fixed values of the lepton-number chemical potential,
$\mu_{L_e}=200$~MeV and $\mu_{L_e}=400$~MeV, respectively. For a fixed
value of $\mu_{L_e}$, we find that the lepton fraction changes only 
slightly with temperature. This is concluded from the comparison of 
the results at $T=0$ (solid lines), $T=20$~MeV (dashed lines), and 
$T=40$~MeV (dotted lines) in Fig.~\ref{plot_Y_0-20-40}.

As is easy to check, at $T=40$~MeV, i.e., when Cooper 
pairing is not so strong, the $\mu$ dependence of $Y_{L_e}$ 
does not differ very much from the prediction in the simple 
two-flavor model in Sec.~\ref{simple}. By saying this, of course, 
one should not undermine the fact that the lepton fraction in 
Fig.~\ref{plot_Y_0-20-40} has a visible structure in the 
dependence on $\mu$ at $T=0$ and $T=20$~MeV. This indicates 
that quark Cooper pairing plays a nontrivial role in determining 
the value of $Y_{L_e}$. 

Our numerical study shows that it is hard to achieve values 
of the lepton fraction more than about $0.05$ in the CFL 
phase. Gapless versions of the CFL phases, on the other 
hand, could accommodate a lepton fraction up to about 
$0.2$ or so, provided the quark and lepton-number chemical 
potentials are sufficiently large. 

{From} Fig.~\ref{plot_Y_0-20-40}, we can also see that the 
value of the lepton fraction $Y_{L_e} \approx 0.4$, i.e.,
the value expected at the center of the protoneutron star 
right after its creation, requires the lepton-number chemical 
potential $\mu_{L_e}$ in the range somewhere between 
$200$~MeV and $400$~MeV, or slightly higher. The larger
the quark chemical potential $\mu$ the larger $\mu_{L_e}$
is needed. Then, in a realistic construction of a star,  
this is likely to result in a noticeable gradient of
the lepton-number chemical potential at the initial time. 
This gradient may play an important role in the subsequent
deleptonization due to neutrino diffusion through dense 
matter.

\subsection{$T$--$\mu_{L_e}$ phase diagram}

Now let us explore the phase diagram in the plane of temperature 
and lepton-number chemical potential, keeping the quark chemical 
potential fixed. Two such slices of the phase diagram are presented 
in Fig.~\ref{phasediagramTnu}. The upper panel corresponds to a  
moderate value of the quark chemical potential, $\mu=400$~MeV. 
This could be loosely termed as the ``outer stellar core'' phase 
diagram. The lower panel in Fig.~\ref{phasediagramTnu} corresponds 
to $\mu=500$~MeV, and we could associate it with the ``inner 
stellar core'' case. Note, however, that the terms ``inner core'' 
and ``outer core'' should not be interpreted literally here. The 
central densities of (proto-)neutron stars are subject to large 
theoretical uncertainties and, thus, are not known very well. 
In the model at hand, the case $\mu = 400$ MeV (``outer core") 
corresponds to a range of densities around $4\rho_0$, while the 
case $\mu = 500$ MeV (``inner core") corresponds to a range of 
densities around $10\rho_0$. These values are of the same order 
of magnitude that one typically obtains in models (see, e.g., 
Ref.~\cite{prakash-et-al}).

In addition to the phase diagrams, we also show the results for the 
dynamical quark masses, the gap parameters, and the charge chemical 
potentials. These are plotted as functions of temperature in 
Fig.~\ref{plot_nu0-200-400_mu400} 
($\mu=400$~MeV, ``outer core'' case) and in 
Fig.~\ref{plot_nu0-200-400_mu500} 
($\mu=500$~MeV, ``inner core'' case), 
for three different values of the lepton-number chemical potential,
$\mu_{L_e}=0$~MeV (upper panels), 
$\mu_{L_e}=200$~MeV (middle panels) 
and $\mu_{L_e}=400$~MeV (lower panels).

At first sight, the two diagrams in Fig.~\ref{phasediagramTnu} 
look so different that no obvious connection between them could 
be made. It is natural to ask, therefore, how such a dramatic 
change could happen with increasing the value of the quark chemical 
potential from $\mu=400$~MeV to $\mu=500$~MeV. In order to understand 
this, it is useful to place the corresponding slices of the phase 
diagram in the three-dimensional diagram in Fig.~\ref{phase3d}. 

The $\mu=500$~MeV diagram corresponds to the right-hand side 
surface of the bounding box in Fig.~\ref{phase3d}. This 
contains almost all complicated phases with strange-nonstrange 
cross-flavor pairing. The $\mu=400$~MeV diagram, on the other 
hand, is obtained by cutting the three-dimensional diagram 
with a plane parallel to the bounding surface, but going 
through the middle of the diagram. This part of the diagram 
is dominated by the 2SC and NQ phases. Keeping in mind the 
general structure of the three-dimensional phase diagram, 
it is also not difficult to understand how the two diagrams 
in Fig.~\ref{phasediagramTnu} transform into each other.

Several comments are in order regarding the zero-temperature
phase transition from the CFL to gCFL$^{\prime}$ phase, shown  
by a small black triangle in the phase diagram at $\mu=500$~MeV, 
see the lower panel in Fig.~\ref{phasediagramTnu}. The appearance 
of this transition is in agreement with the analytical result in 
Sec.~\ref{simple}. Moreover, the critical value of the lepton-number
chemical potential also turns out to be very close to the estimate
in Eq.~(\ref{mu_L^cr}). Indeed, by taking into account that 
$M_{s}\approx 214$~MeV and $\Delta_2\approx 76$~MeV, we obtain 
$\mu_{L_e}^{\rm (cr)}=\Delta_2-M_s^2/(2\mu) \approx 30$~MeV 
which agrees well with the numerical value. 

In order to estimate how the critical value of $\mu_{L_e}$ changes 
with decreasing the quark chemical potential below $\mu=500$~MeV, 
we can use the zero-temperature numerical results for $M_{s}$ and 
$\Delta_2$ from Ref.~\cite{pd-mass}. Then, we arrive at the following 
power-law fit for the $\mu$-dependence of the critical value:
\be
\mu_{L_e}^{\rm (cr)} \approx \frac{23}{4000}\left(\mu-457.4\right)
\left(622.1-\mu\right), 
\label{mu_l^cr-fit}
\ee
for $457.4 \leq \mu \leq 500$~MeV (in the fit, $\mu_{L_e}^{\rm (cr)}$ 
and $\mu$ are measured in MeV). Note that the CFL phase does 
not appear at $T=0$ when $\mu<457.4$~MeV \cite{pd-mass}. With 
increasing the values of the quark chemical potential above 
$\mu=500$~MeV, one expects that the critical value of $\mu_{L_e}$ 
should continue to increase for a while, and then decrease when 
the effects of the cutoff start to suppress the size of the gap 
$\Delta_2$. However, the validity of the fit in Eq.~(\ref{mu_l^cr-fit}) 
is questionable there because no numerical data for $\mu>500$~MeV 
was used in its derivation.

Before concluding this subsection, we should mention that a 
schematic version of the phase diagram in $T$--$\mu_{L_e}$ plane
was earlier presented in Ref.~\cite{SRP}, see the left panel in
Fig.~4 there. In Ref.~\cite{SRP}, the value of the quark 
chemical potential was $\mu=460$~MeV, and therefore a direct 
comparison with our results is not simple. One can see, 
however, that the diagram of Ref.~\cite{SRP} fits naturally 
into the three-dimensional diagram in Fig.~\ref{phase3d}. 
Also, the diagram of Ref.~\cite{SRP} is topologically close
to our version of the phase diagram at $\mu=500$~MeV, shown 
in the lower panel of Fig.~\ref{phasediagramTnu}. The 
quantitative difference is not surprising: the region of the 
(g)CFL phase is considerably larger at $\mu=500$~MeV than 
at $\mu=460$~MeV.

\section{Conclusions}
\label{conclusions}

\begin{figure*}[h!]
  \begin{center}
    \includegraphics[width=0.85\textwidth]{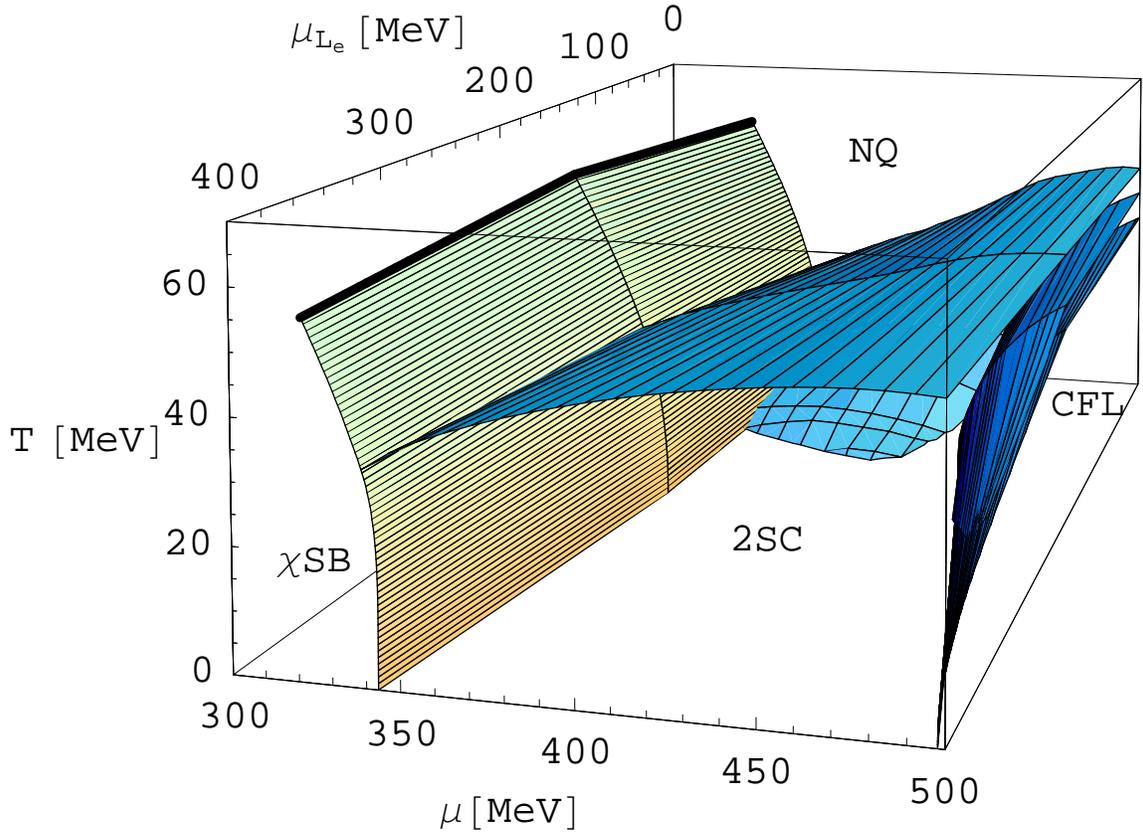}
    \caption{General structure of the phase diagram of neutral dense 
     quark matter in the three-dimensional space spanned by the quark 
     chemical potential $\mu$, the lepton-number chemical potential $\mu_{L_e}$,
     and the temperature $T$.}
    \label{phase3d}
  \end{center}
\end{figure*}

\begin{figure*}[h!]
  \begin{center}
    \includegraphics[width=0.6\textwidth]{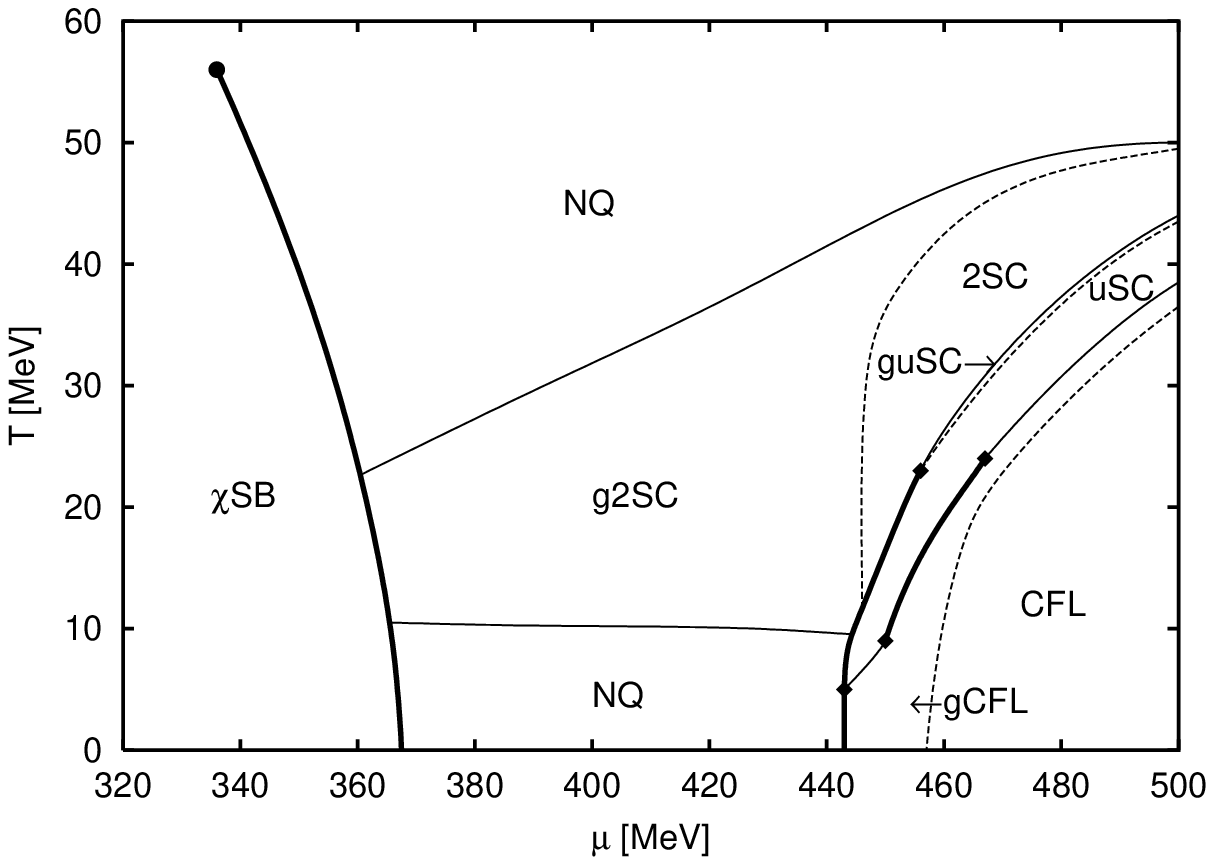}\\
    \includegraphics[width=0.6\textwidth]{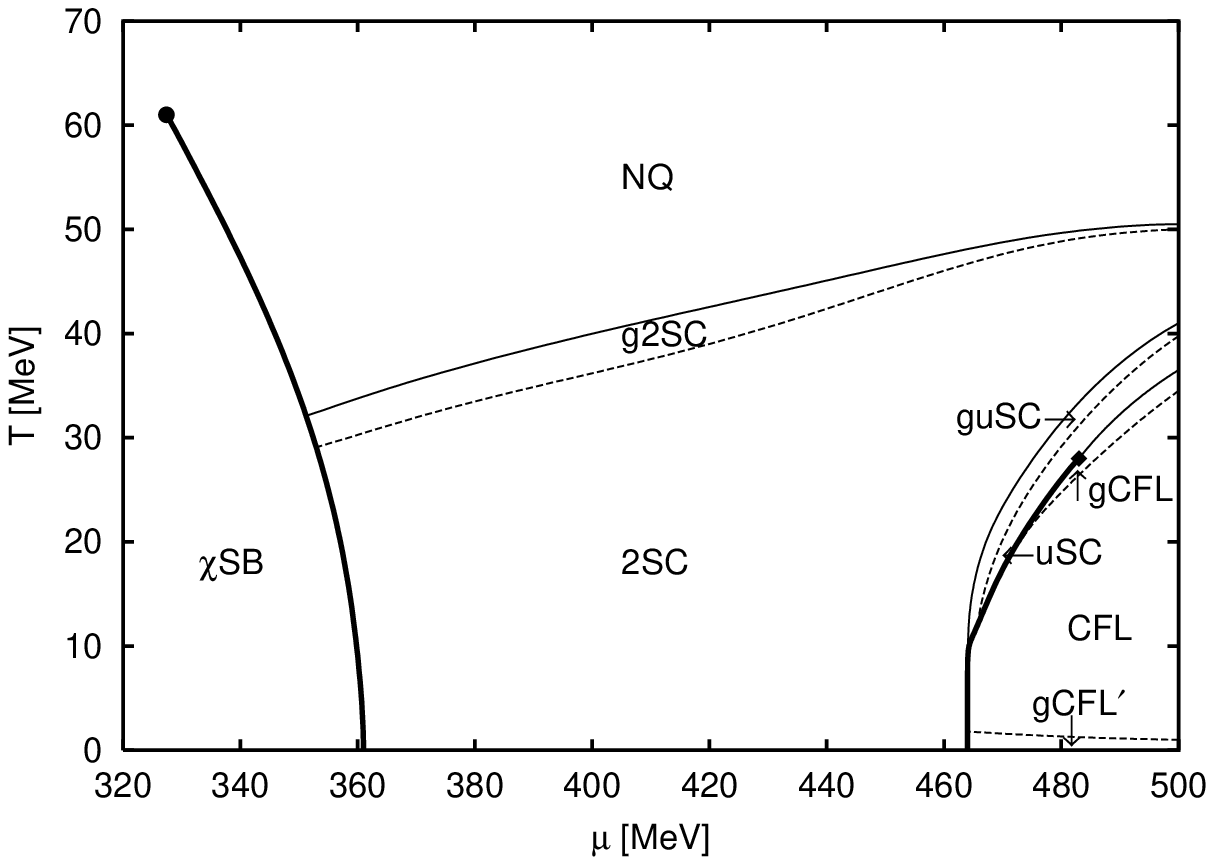}\\
    \includegraphics[width=0.6\textwidth]{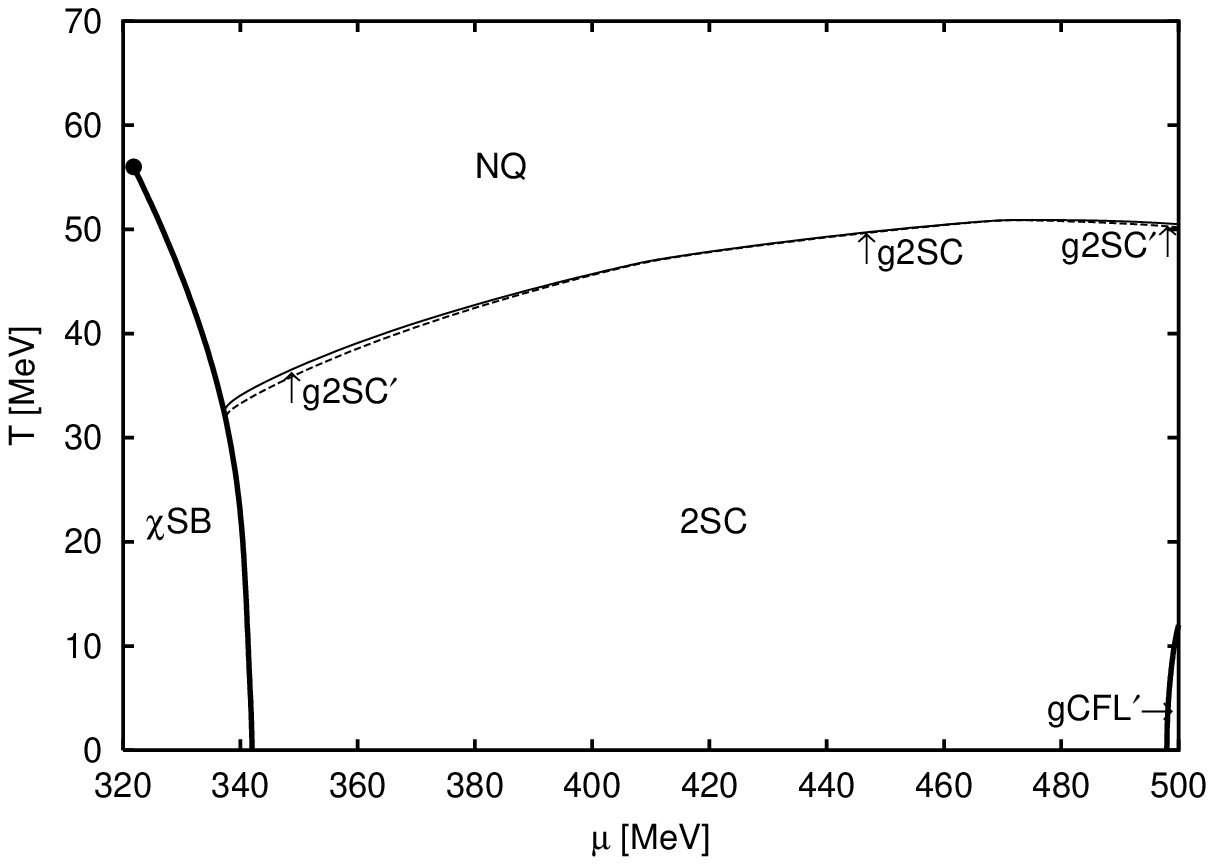}
    \caption{The phase diagrams of neutral quark matter at fixed
     lepton-number chemical potentials $\mu_{L_e}=0$ (upper panel), 
     $\mu_{L_e}=200$~MeV (middle panel), and $\mu_{L_e}=400$~MeV 
     (lower panel). Note that the upper diagram, describing matter 
     in absence of neutrino trapping, is the same as in Fig.~1 of 
     Ref.~\cite{pd-mass}.}
    \label{phasediagram_mu-T}
  \end{center}
\end{figure*}

\begin{figure*}[h!]
  \begin{center}
    \includegraphics[width=0.9\textwidth]{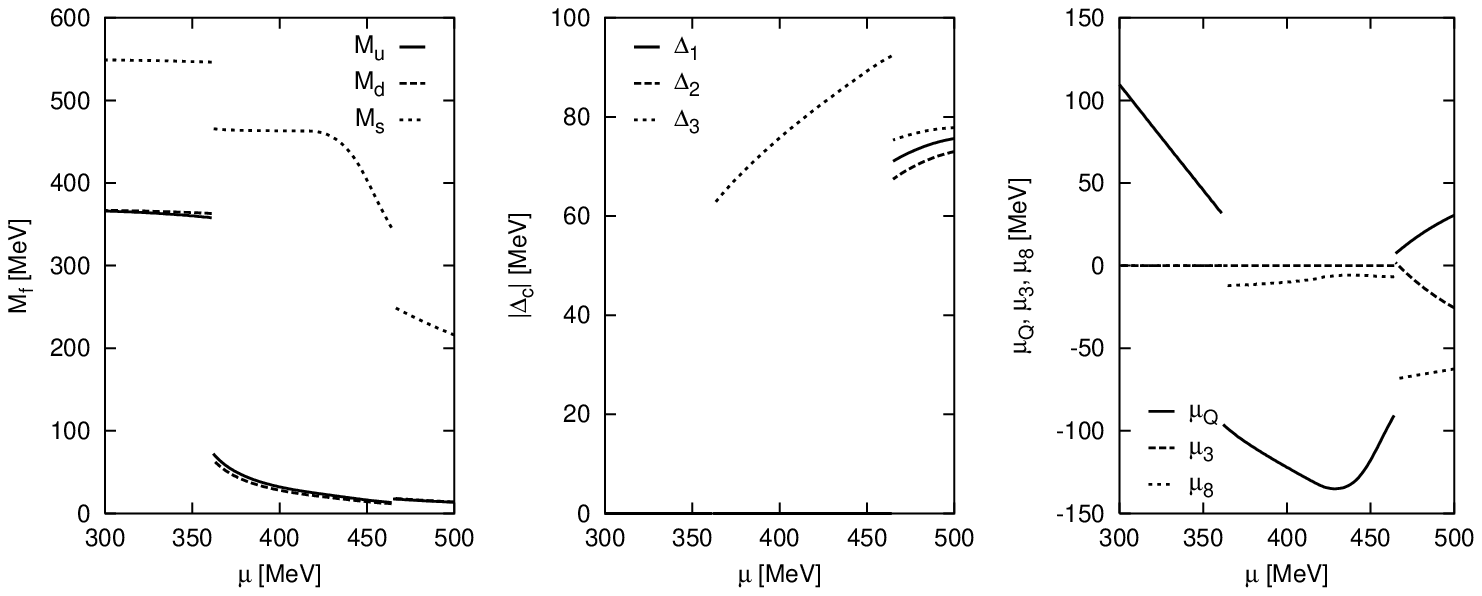}\\
    \includegraphics[width=0.9\textwidth]{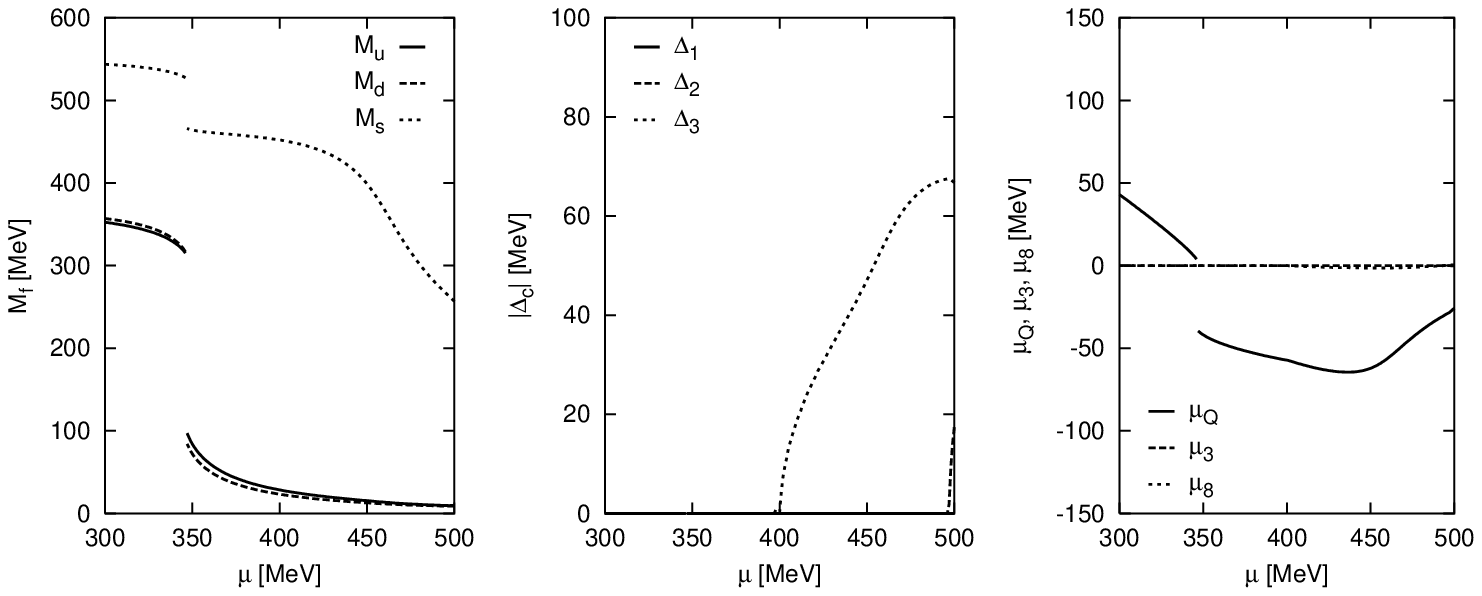}
    \caption{Dependence of the quark masses, of the gap parameters,
     and of the electric and color charge chemical potentials on the
     quark chemical potential at a fixed temperature, $T=0~\mbox{MeV}$ 
     (three upper panels) 
     and $T=40~\mbox{MeV}$ (three lower panels). The lepton-number 
     chemical potential is kept fixed at $\mu_{L_e}=200$~MeV.}
    \label{plot0-20-40_200}
  \end{center}
\end{figure*}

\begin{figure*}[h!]
  \begin{center}
    \includegraphics[width=0.9\textwidth]{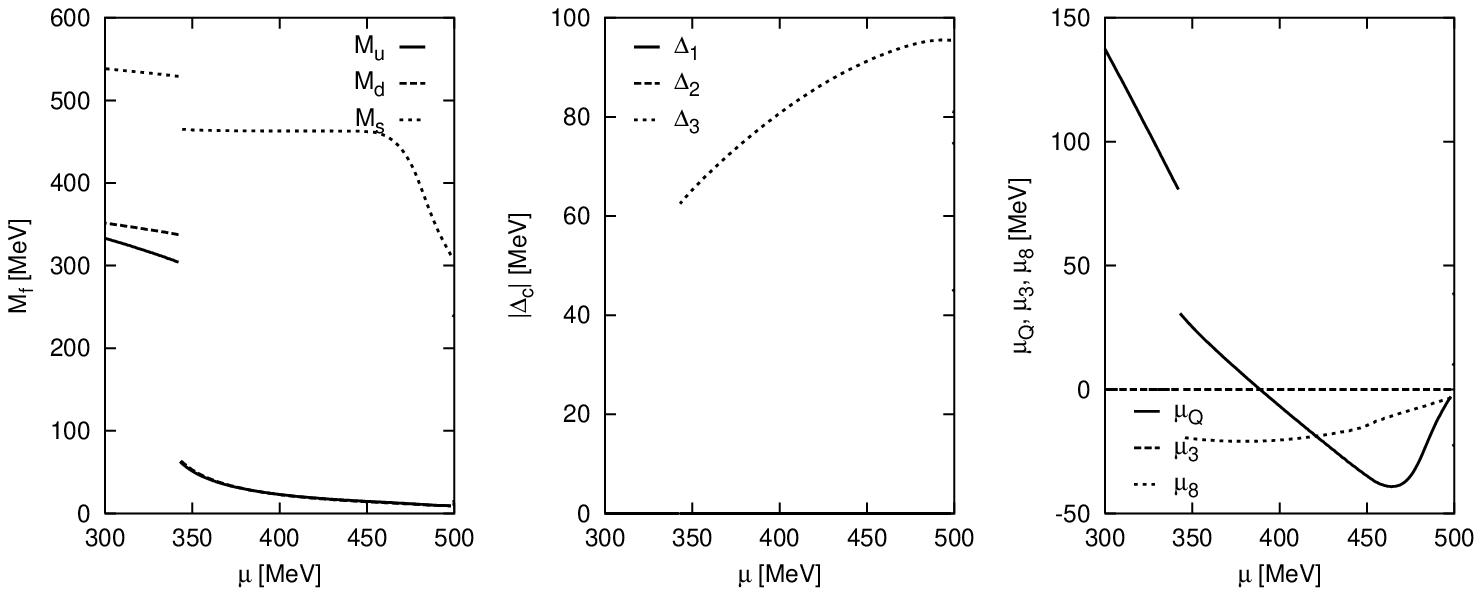}\\
    \includegraphics[width=0.9\textwidth]{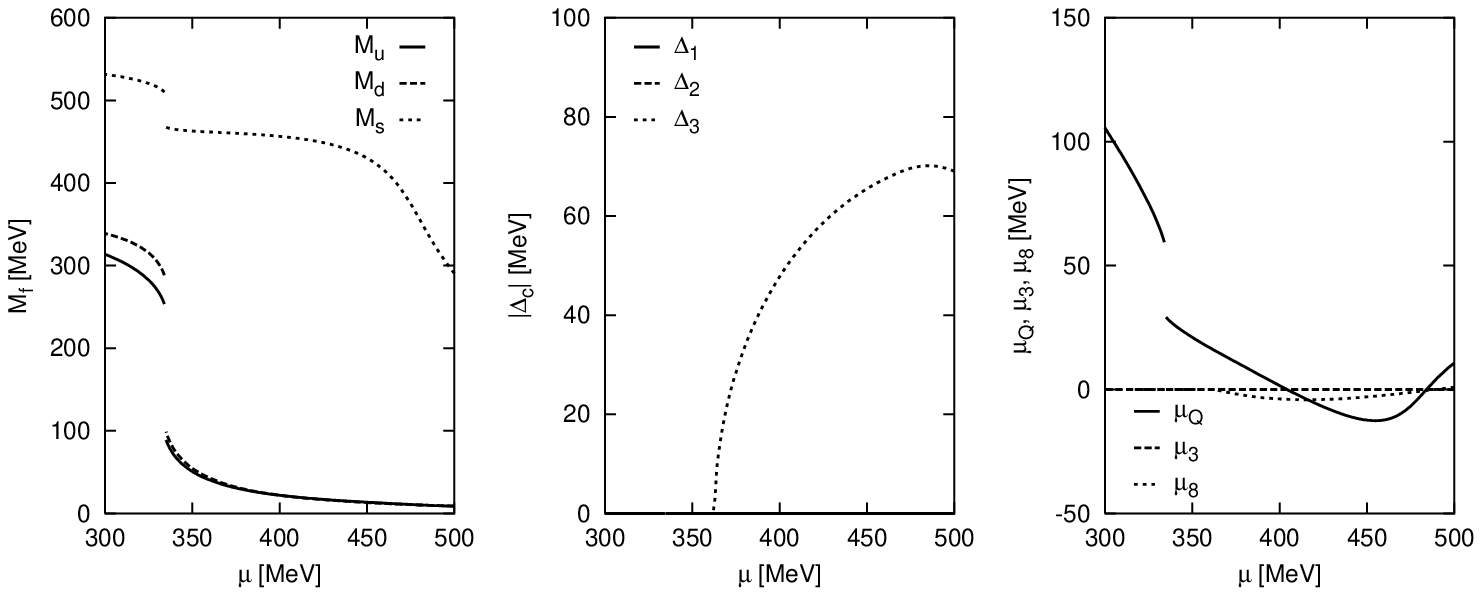}
    \caption{Dependence of the quark masses, of the gap parameters,
     and of the electric and color charge chemical potentials on the
     quark chemical potential at a fixed temperature, 
     $T=0~\mbox{MeV}$ (three upper panels), and 
     $T=40~\mbox{MeV}$ (three lower panels). The lepton-number 
     chemical potential is kept fixed at $\mu_{L_e}=400$~MeV.}
    \label{plot0-20-40_400}
  \end{center}
\end{figure*}

\begin{figure*}[h!]
  \begin{center}
    \includegraphics[angle=-90,width=0.51\textwidth]{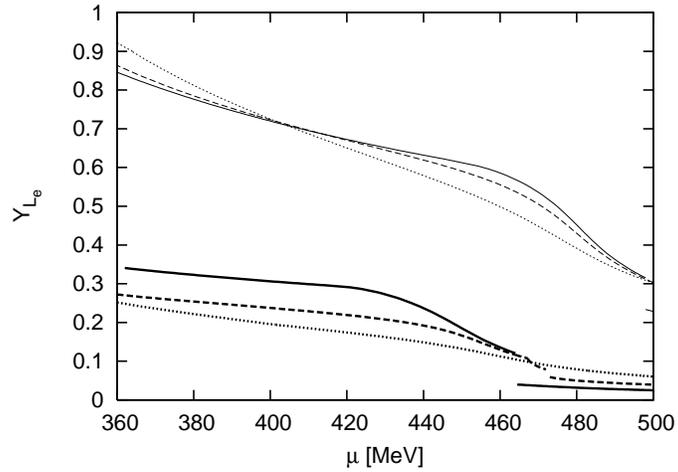}
    \caption{Dependence of the electron family lepton fraction $Y_{L_e}$
     for $\mu_{L_e} = 200$~MeV (thick lines) and $\mu_{L_e} = 400$~MeV
     (thin lines) on the quark chemical potential at a fixed temperature, 
     $T=0~\mbox{MeV}$ (solid lines), $T=20~\mbox{MeV}$ (dashed lines), and 
     $T=40~\mbox{MeV}$ (dotted lines).}
    \label{plot_Y_0-20-40}
  \end{center}
\end{figure*}

\begin{figure*}[ht]
  \begin{center}
    \includegraphics[width=0.75\textwidth]{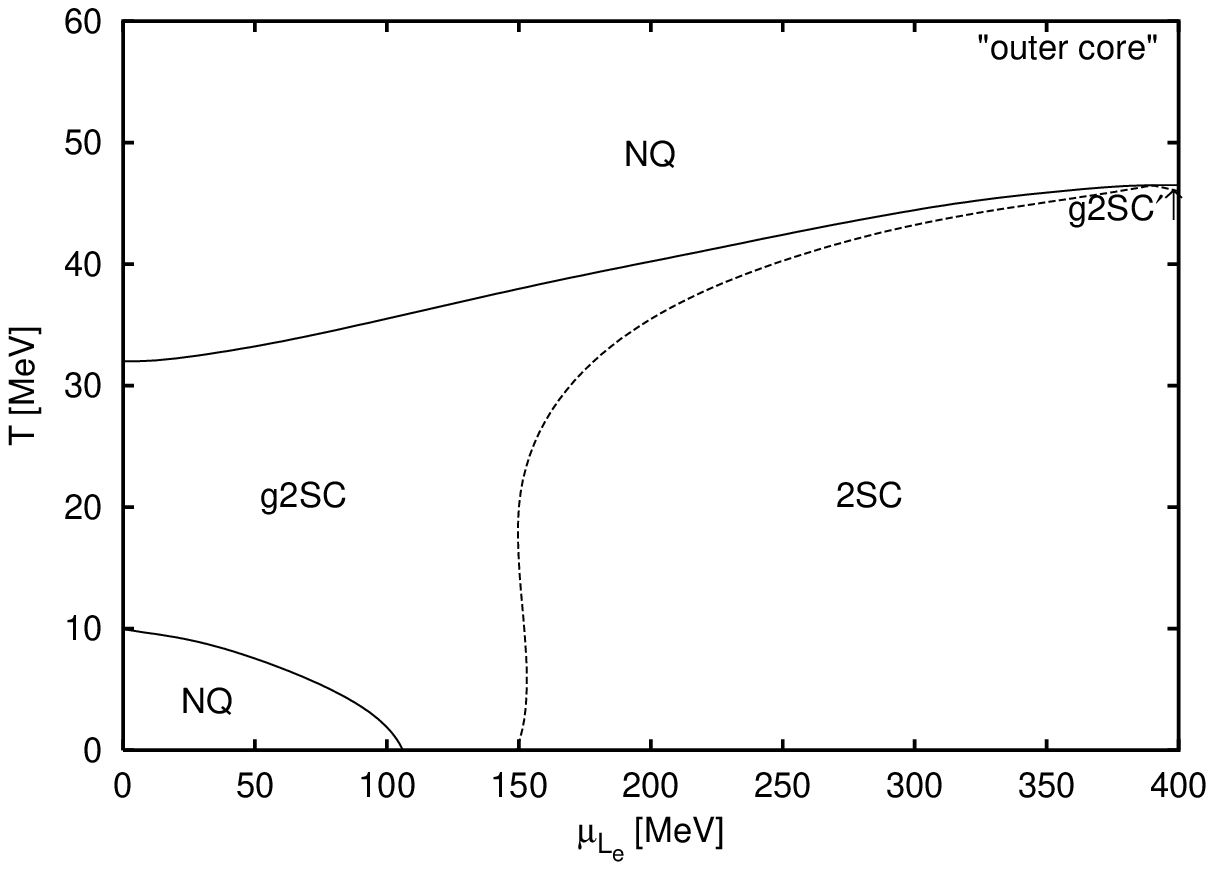}\\
    \includegraphics[width=0.75\textwidth]{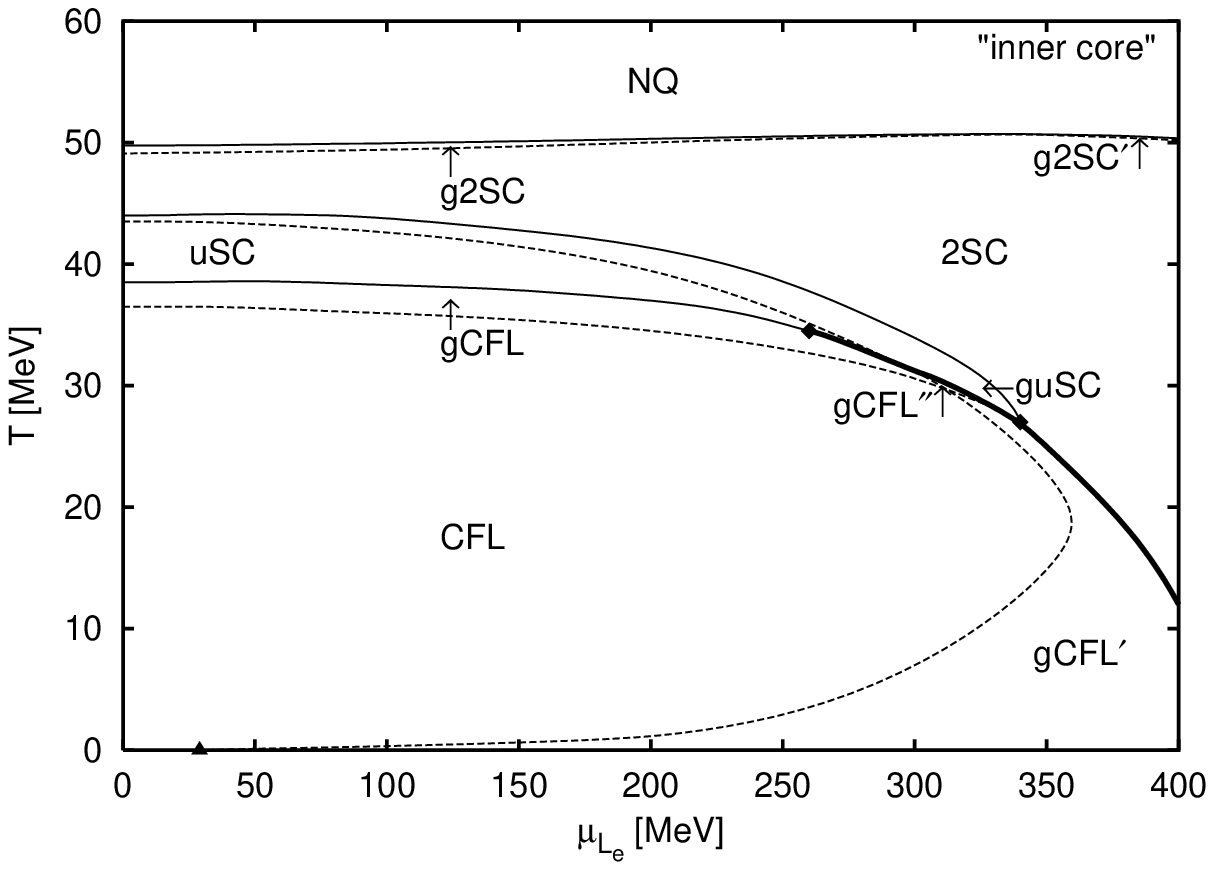}
    \caption{The phase diagrams of neutral quark matter in the plane of
      temperature and lepton-number chemical potential at two fixed
      values of quark chemical potential: $\mu=400$~MeV (upper panel)
      and $\mu=500$~MeV (lower panel). The triangle in the lower panel 
      denotes the transition point from the CFL phase to the
      gCFL$^{\prime}$ phase at $T=0$.}
  \label{phasediagramTnu}
  \end{center}
\end{figure*}

\begin{figure*}[ht]
  \begin{center}
    \includegraphics[width=0.95\textwidth]{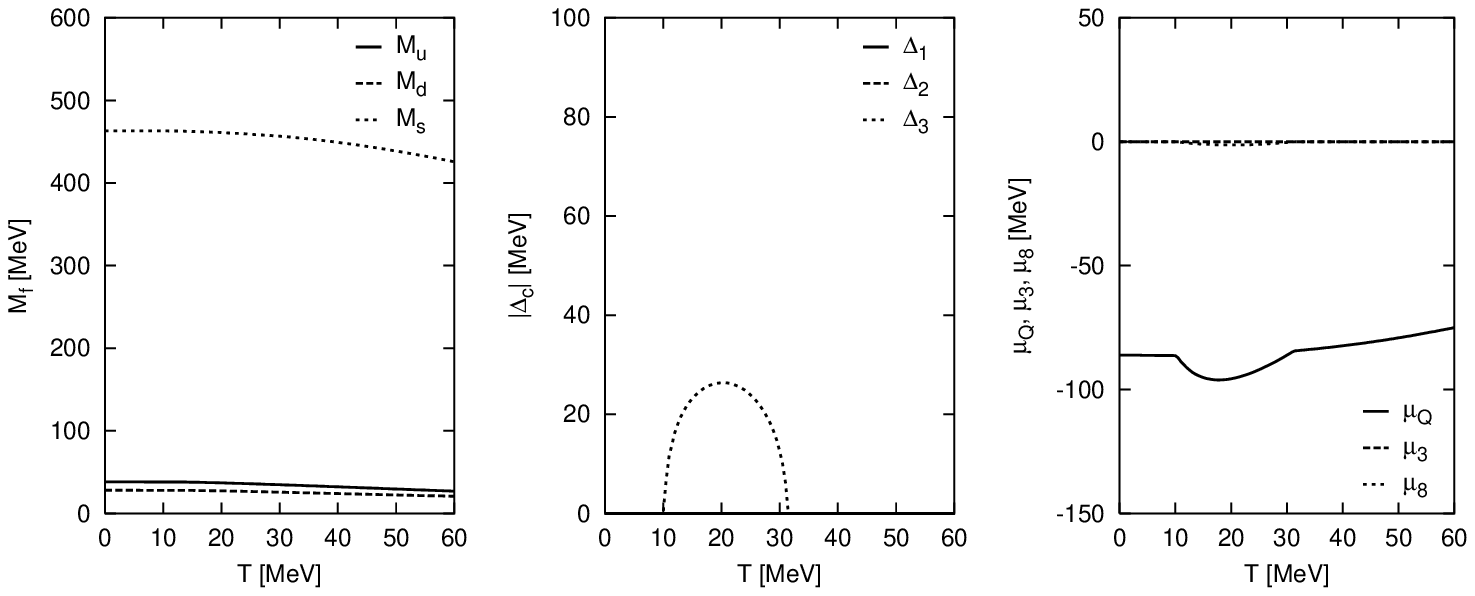}\\
    \includegraphics[width=0.95\textwidth]{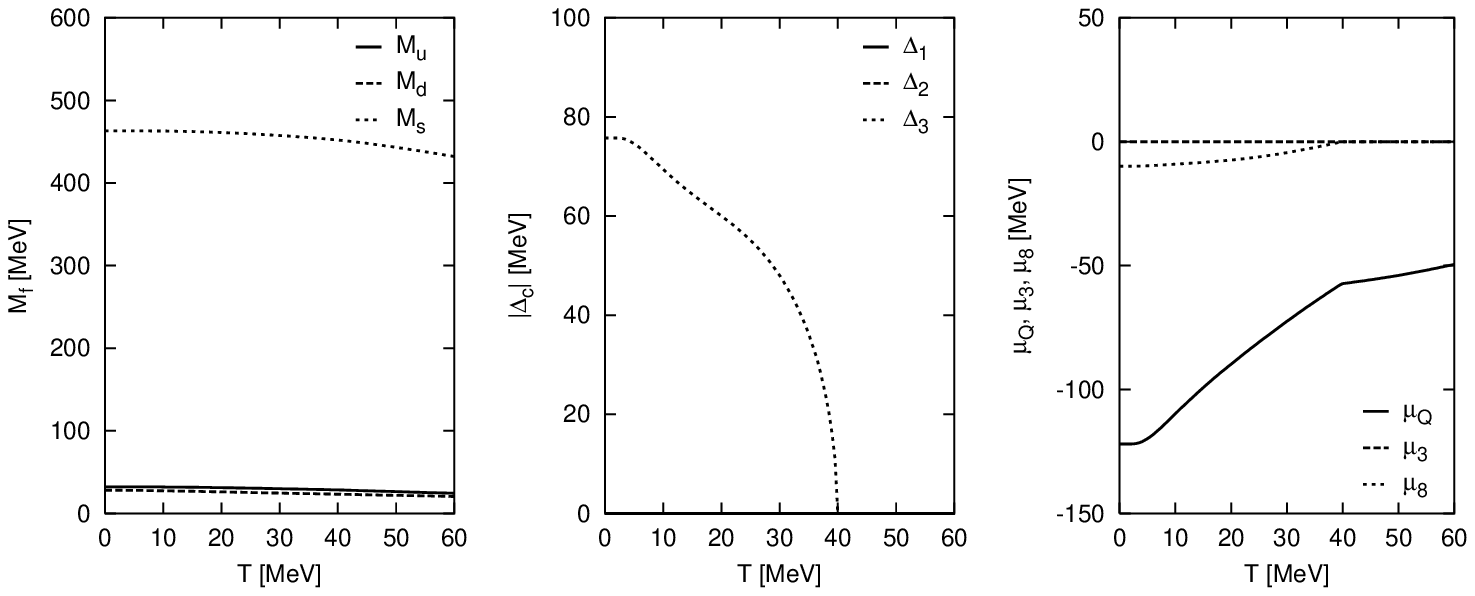}\\
    \includegraphics[width=0.95\textwidth]{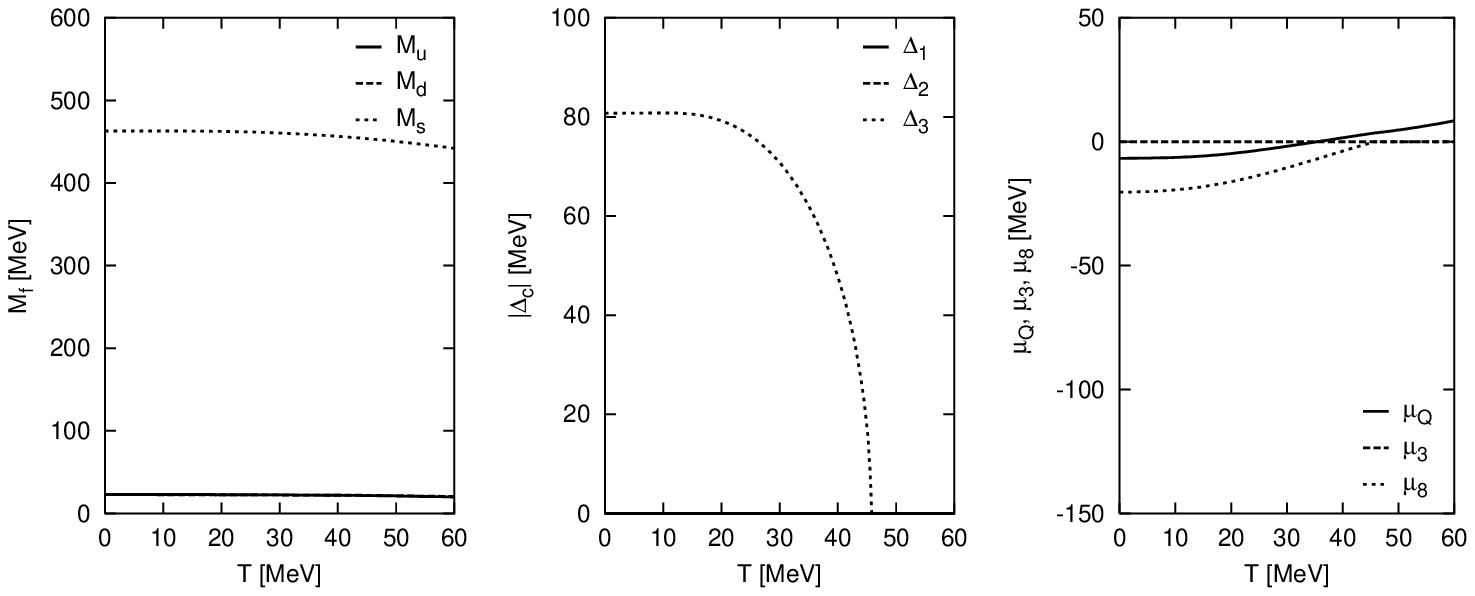}
    \caption{Dependence of the quark masses, of the gap parameters,
     and of the electric and color charge chemical potentials on the
     temperature at a fixed value of the lepton-number chemical potential, 
     $\mu_{L_e}=0~\mbox{MeV}$ (three upper panels), 
     $\mu_{L_e}=200~\mbox{MeV}$ (three middle panels), and 
     $\mu_{L_e}=400~\mbox{MeV}$ (three lower panels). 
     The quark chemical potential is $\mu=400$~MeV which may correspond  
     to the conditions in the outer quark core inside a neutron star.}
    \label{plot_nu0-200-400_mu400}
  \end{center}
\end{figure*}

\begin{figure*}[ht]
  \begin{center}
    \includegraphics[width=0.95\textwidth]{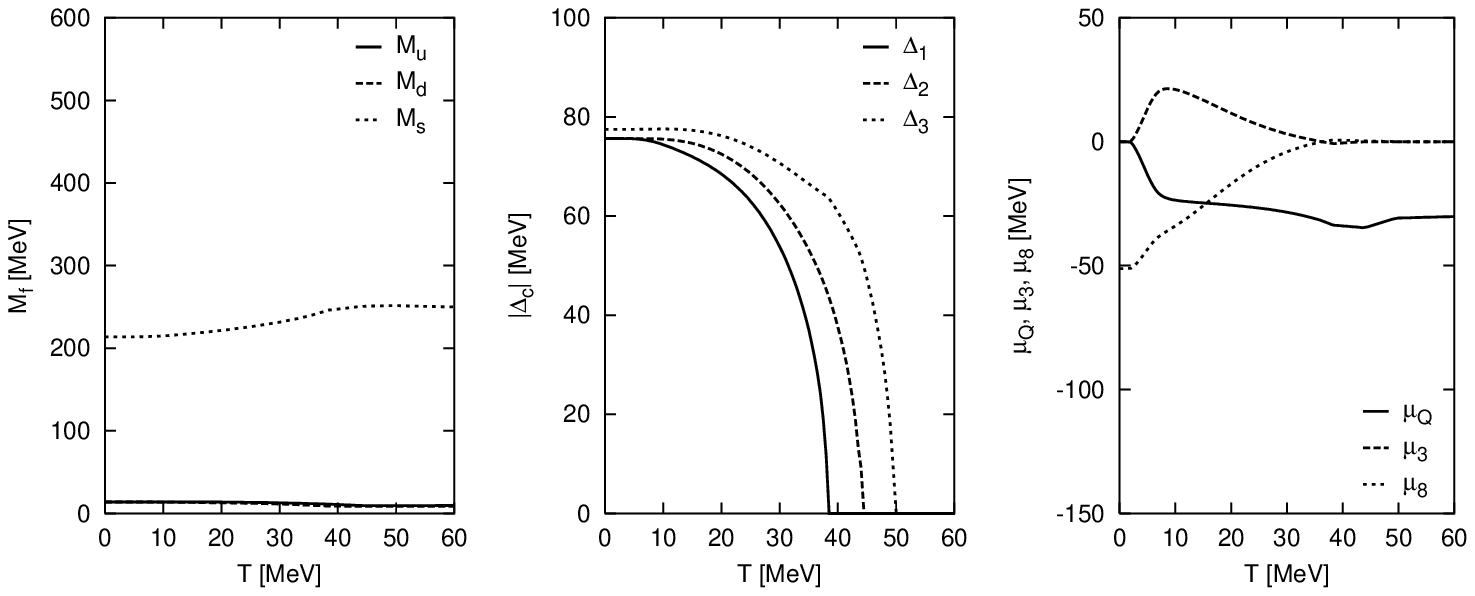}\\
    \includegraphics[width=0.95\textwidth]{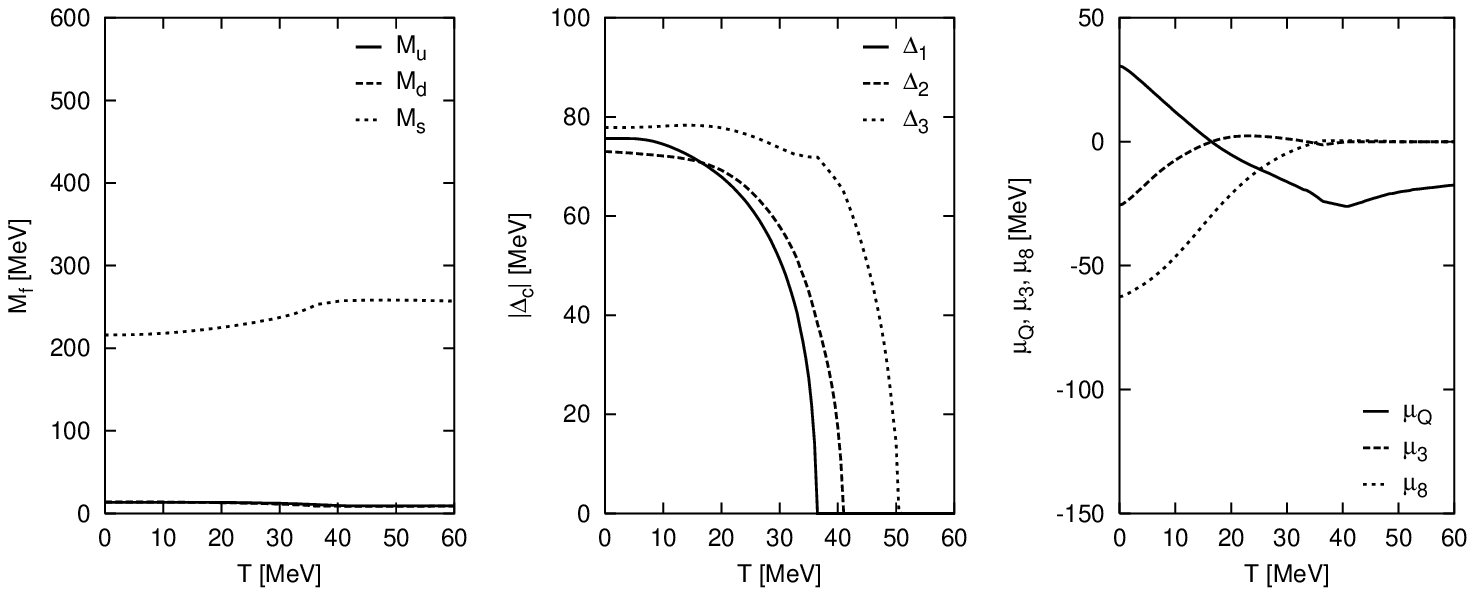}\\
    \includegraphics[width=0.95\textwidth]{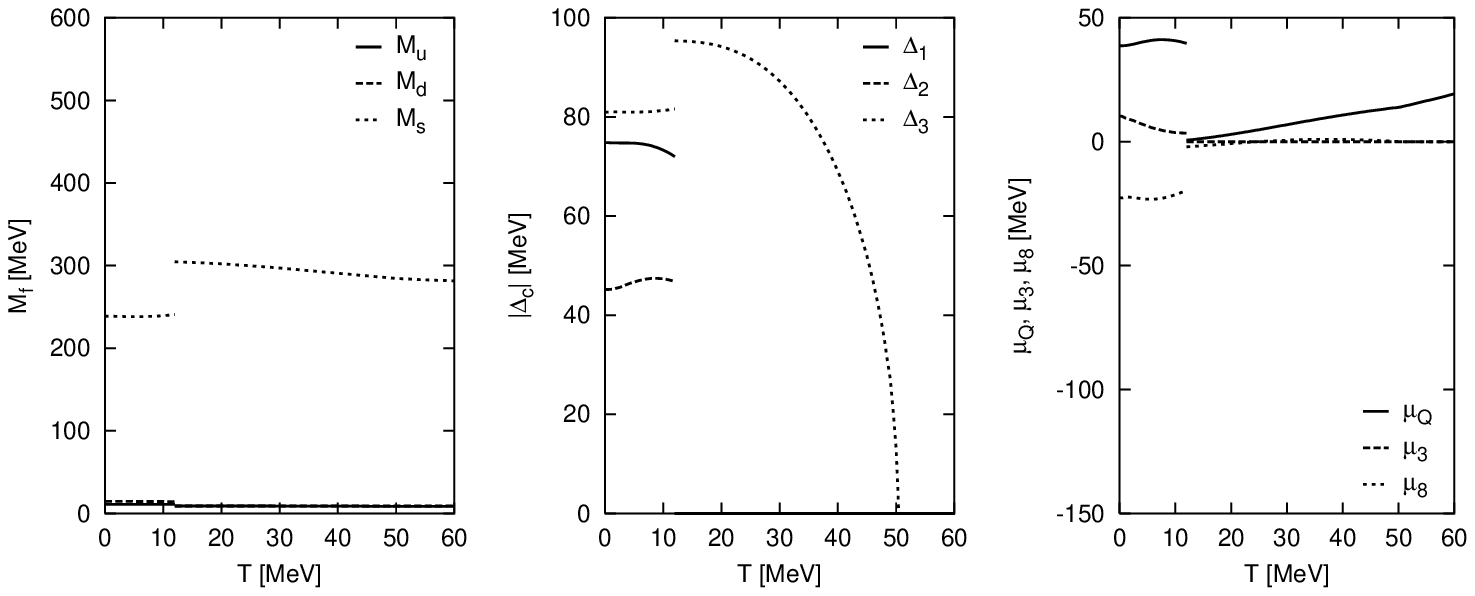}
    \caption{Dependence of the quark masses, of the gap parameters,
     and of the electric and color charge chemical potentials on the
     temperature at a fixed value of the lepton-number chemical potential, 
     $\mu_{L_e}=0~\mbox{MeV}$ (three upper panels), 
     $\mu_{L_e}=200~\mbox{MeV}$ (three middle panels), and 
     $\mu_{L_e}=400~\mbox{MeV}$ (three lower panels). 
     The quark chemical potential is $\mu=500$~MeV which may correspond  
     to the conditions at the center of a quark core inside a neutron star.}
    \label{plot_nu0-200-400_mu500}
  \end{center}
\end{figure*}

In this paper, we studied the effect of neutrino trapping on
the phase diagram of neutral three-flavor quark matter within 
the NJL model of Ref.~\cite{RKH}. The results are obtained in 
the mean-field approximation, treating constituent quark masses
as dynamically generated quantities. The overall structure of 
the phase diagram in the space of three parameters, namely
temperature $T$, quark chemical potential $\mu$, and 
lepton-number chemical potential $\mu_{L_{e}}$, is summarized 
in Fig.~\ref{phase3d}. This is further detailed in several 
two-dimensional slices of the phase diagram, including diagrams 
in the plane of temperature and quark chemical potential (see 
Fig.~\ref{phasediagram_mu-T}) and in the plane of temperature and 
lepton-number chemical potential (see Fig.~\ref{phasediagramTnu}).

By making use of simple model-independent arguments, as well as 
detailed numerical calculations in the framework of an NJL-type
model, we find that neutrino trapping helps Cooper pairing in
the 2SC phase and suppresses the CFL phase. In essence, this
is the consequence of satisfying the electric neutrality 
constraint in the quark system. In two-flavor quark matter, 
the (positive) lepton-number chemical potential $\mu_{L_e}$ 
helps to provide extra electrons without inducing a large 
mismatch between the Fermi momenta of up and down quarks. 
With reducing the mismatch, of course, Cooper pairing gets 
stronger. This is in sharp contrast to the situation in the
CFL phase of quark matter, which is neutral in the absence 
of electrons. Additional electrons due to large $\mu_{L_e}$ 
can only put extra stress on the system.

In application to protoneutron stars, our findings in this paper
strongly suggest that the CFL phase is very unlikely to appear
during the early stage of the stellar evolution before the 
deleptonization is completed. If color superconductivity occurs
there, the 2SC phase is the best candidate for the ground state.
In view of this, it might be quite natural to suggest 
that matter inside protoneutron stars contains little or no 
strangeness (just as the cores of the progenitor stars) during 
the early times of their evolution. In this connection, it is 
appropriate to recall that neutrino trapping also suppresses 
the appearance of strangeness in the form of hyperonic matter 
and kaon condensation \cite{prakash-et-al}. Our results, therefore, 
are a special case of a generic property.

After the deleptonization occurs, it is possible that the ground 
state of matter at high density in the central region of the star 
is the CFL phase. This phase contains a large number of strange
quarks. Therefore, an abundant production of strangeness should
happen right after the deleptonization in the protoneutron star. 
If realized in nature, in principle this scenario may have 
observational signatures.

\begin{acknowledgments}
This work was supported in part by the Virtual Institute of the
Helmholtz Association under grant No. VH-VI-041 and by Gesellschaft
f\"{u}r Schwerionenforschung (GSI) and by Bundesministerium f\"{u}r
Bildung und Forschung (BMBF). S.~R. thanks for using the Center
for Scientific Computing (CSC) of the Johann Wolfgang
Goethe-Universit\"at Frankfurt am Main.
\end{acknowledgments}




\end{document}